\documentclass[11pt]{article}
\usepackage[utf8]{inputenc}
\usepackage{geometry}
\newcommand{\citeg}[1]{\citep[e.g.,][]{#1}}
\newcommand{\add}[1]{\textcolor{black}{#1}} 
\newcommand{\addd}[1]{\textcolor{black}{#1}}

\usepackage{authblk}
\usepackage{natbib}
\usepackage{color}
\usepackage{amsmath}
\usepackage{amssymb}
\usepackage{graphicx}
\usepackage[dvipsnames]{xcolor}
\usepackage[font={small,it}]{caption}
\usepackage{lineno}
\usepackage{epigraph}
\usepackage{soul}

\title{Mercury, Moon, Mars: Surface expressions of mantle convection and interior evolution of stagnant-lid bodies}
\author[1,2]{Nicola Tosi}
\author[1]{Sebastiano Padovan}
\affil[1]{Deutsches Zentrum f\"{u}r Luft- und Raumfahrt (DLR), Institute of Planetary Research, Berlin, Germany}
\affil[2]{Technische Universit\"{a}t, Department of Astronomy and Astrophysics, Berlin, Germany}
\date{}                     
\begin{document}

\thispagestyle{empty}
\vspace*{5cm}
\begin{center}
\LARGE{Mercury, Moon, Mars: Surface expressions of mantle convection and interior evolution of stagnant-lid bodies}

\vspace{0.5cm}

\large{Nicola Tosi$^{1,2}$ and Sebastiano Padovan$^{1}$}

\vspace{0.5cm}
\small{$^{1}$\textit{Deutsches Zentrum f\"{u}r Luft- und Raumfahrt (DLR), Institute of Planetary Research, Berlin, Germany}} \\
\small{$^{2}$ \textit{Technische Universit\"{a}t, Department of Astronomy and Astrophysics, Berlin, Germany}}

\vspace{2cm}
\large{Accepted chapter to appear in}

\vspace{0.5cm}
\large{Mantle Convection and Surface Expressions\\ H. Marquardt, M. Ballmer, S. Cottar, K. Jasper (eds.)\\ \textit{AGU Monograph Series}, 2020.}

\end{center}

\newpage
\pagenumbering{arabic}
\setcounter{page}{1}

\maketitle

\epigraph{It is hard to be finite upon an infinite subject, and all subjects are infinite}{\textit{Herman Melville}}

\begin{center}
    \textbf{Abstract}
\end{center}

\add{The evolution of the interior of stagnant-lid bodies is comparatively easier to model and predict with respect to the Earth's due to the absence of the large uncertainties associated with the physics of plate tectonics, its onset time and efficiency over the planet's history.}
Yet, the observational record for these bodies is both scarcer and sparser with respect to the Earth's. It is restricted to a limited number of samples and a variety of remote-sensing measurements of billions-of-years-old surfaces whose actual age is difficult to determine precisely. Combining these observations into a coherent picture of the thermal and convective evolution of the planetary interior represents thus a major challenge. In this chapter, we review key processes and (mostly geophysical) observational constraints that can be used to infer the global characteristics of mantle convection and thermal evolution of the interior of Mercury, the Moon and Mars, the three major terrestrial bodies where a stagnant lid has likely been present throughout most of their history. 

\section{Introduction}

In contrast to the Earth whose evolution is controlled by plate tectonics and surface recycling, the other rocky bodies of the solar system operate today in a stagnant-lid mode. \add{While Venus has and had a complex and rather poorly understood tectonics,} Mercury, the Moon, and Mars are the paradigm of stagnant-lid bodies. They possess a single, continuous lithosphere, not fragmented into tectonic plates, which undergoes deformation to a much lower extent than the Earth’s, and below which solid-state mantle convection takes \add{(or took)} place \citeg{Breuer2015}. With the possible exception of Mars, which could have experienced a brief episode of surface mobilization in its earliest history \citep{Nimmo2000,Tosi2013b,Plesa2014}, the three bodies have been characterized by stagnant-lid convection over \add{the largest part of} their evolution as their old and highly cratered surfaces testify. 

While Earth's plate tectonics allows for continuous arc, mid-ocean-ridge, and hot-spot volcanism, stagnant-lid bodies experience only the latter form, which causes mantle melting and the production of secondary crust to be largely concentrated during the early evolution and to fade rapidly as the mantle begins to cool and the stagnant lid to thicken. 
Sinking of cold tectonic plates on Earth causes efficient cooling of the mantle and core. It promotes 
core convection and associated solidification of an inner core, resulting in the generation of a magnetic field. The presence of a stagnant lid tends to prevent heat loss, maintaining the interior warm with fundamental consequences for the mode and timing of magnetic field generation. 

The absence of a recycling mechanism in stagnant-lid bodies makes the surface record of billions of years of evolution shaped by impacts and volcanism available to remote or in-situ inspection \citeg{Fassett2016}. Measurements by orbiting spacecrafts of gravity, topography, surface composition, and magnetic field, combined with surface imaging and analysis of samples and meteorites (available for the Moon and Mars, but not for Mercury) provide a set of observational constraints that can be combined to investigate the thermal history of the interior and how mantle convection shaped it. 

Venus shows a uniformly young surface \citeg{Hauck1998} as indicated by the random distribution of impact craters \citep{Herrick1994}. It does not show signs of Earth-like plate tectonics. Sites of potential subduction \citeg{Schubert1995}, possibly induced by mantle plumes \citep{Davaille2017}, suggest that Venus is neither in a plate-tectonic nor in a classical stagnant-lid regime. The additional lack of a global magnetic field and the extreme paucity of observational constraints on the interior structure (e.g., Venus' moment of inertia is not known, preventing any meaningful estimate of its core size) strongly limit the possibility to infer the convective evolution of the planet as opposed to pure stagnant-lid bodies such as Mercury, the Moon, and Mars, to which we limit the present work. \add{Besides Venus, we will also omit discussing Jupiter's moon Io. Although strictly speaking Io is also a stagnant-lid body, its interior dynamics \citep[e.g.,][]{Tackley2001,Moore2003} and surface tectonics \citep[e.g.,][]{Bland2016} are largely controlled by tidal dissipation \citep[e.g.,][]{Peale1979,Segatz88}. 
The associated heat production is at the origin of Io's massive volcanic activity, which controls heat loss via the so-called heat piping mechanism \citep[e.g.,][]{Oreilly1981,Moore2013} and sets this body apart from those whose evolution is governed by solid-state convection driven by radiogenic heating and secular cooling.}


The chapter is divided in \add{largely self-contained} sections according  to physical processes and, loosely, on their temporal extent.
Accordingly, Section \ref{Sec__MagmaOceans}  discusses some developments in the understanding  of the earliest stages of the evolution characterized  by the solidification of magma oceans and its  potential consequences for the creation of a primary  crust and the onset of mantle convection. Following magma ocean solidification, the mantle will remove heat through convection and associated decompression melting, which contributes to the creation of a secondary crust, a  topic discussed in Section \ref{Sec__Crust}. Large impacts were common during the first  phases of the solar system, sometimes with important implications for  the overall evolution of the body.  We describe the potential interaction  of the impact-delivered energy with mantle convection in Section \ref{Sec__Impacts}. The evidence for the presence of ancient magnetic fields on all  the three bodies and a present magnetic field on Mercury provides insights into the cooling of the core, as discussed in Section \ref{Sec__MagField}. Section \ref{Sec__AdditionalConstraints} describes  a  set of additional, though somewhat more indirect, constraints on the thermal evolution provided by evidences for global contraction  and expansion, measurements of the heat flux, and inferences on the lithospheric thickness based on analysis of gravity and topography. We conclude in Section \ref{Section__Summary}  with a summary comparing the evolution of Mercury,  the Moon, and Mars in light of the  constraints discussed throughout the chapter.


\section{Magma ocean solidification and onset of solid-state mantle convection}\label{Sec__MagmaOceans}

During the early stages of the solar system, accretionary impacts of planetesimals, high rates of radiogenic heat production, core-mantle differentiation, and giant collisions involving planetary embryos and protoplanets all contribute to store vast amounts of thermal energy into forming bodies. The accompanying temperatures can easily exceed the liquidus of silicates, leading to the formation of magma oceans, which can extend to great depths, possibly over the entire mantle \citep[e.g.,][]{ElkinsTanton12, Solomatov2015}. The crystallization of a magma ocean sets the stage for the subsequent long-term evolution of the planet: it controls the initial silicate differentiation of the mantle and the generation of the first (primordial) crust \citep[e.g.,][]{Wood1970,Warren1985,Elkins-Tanton2005b,VanderKaaden2015}, affects the volatile budget of the interior and atmosphere \citep[e.g.,][]{Abe1986,Elkins-Tanton2008,Lebrun2013,Hier2017,Nikolaou2019}, and determines the earliest forms of mantle convection and surface tectonics \citep{Tosi2013b, Maurice2017, Ballmer2017, Boukare2018, Morison2019}. 

\begin{figure}[ht!]
\centering
\includegraphics[width=1\textwidth]{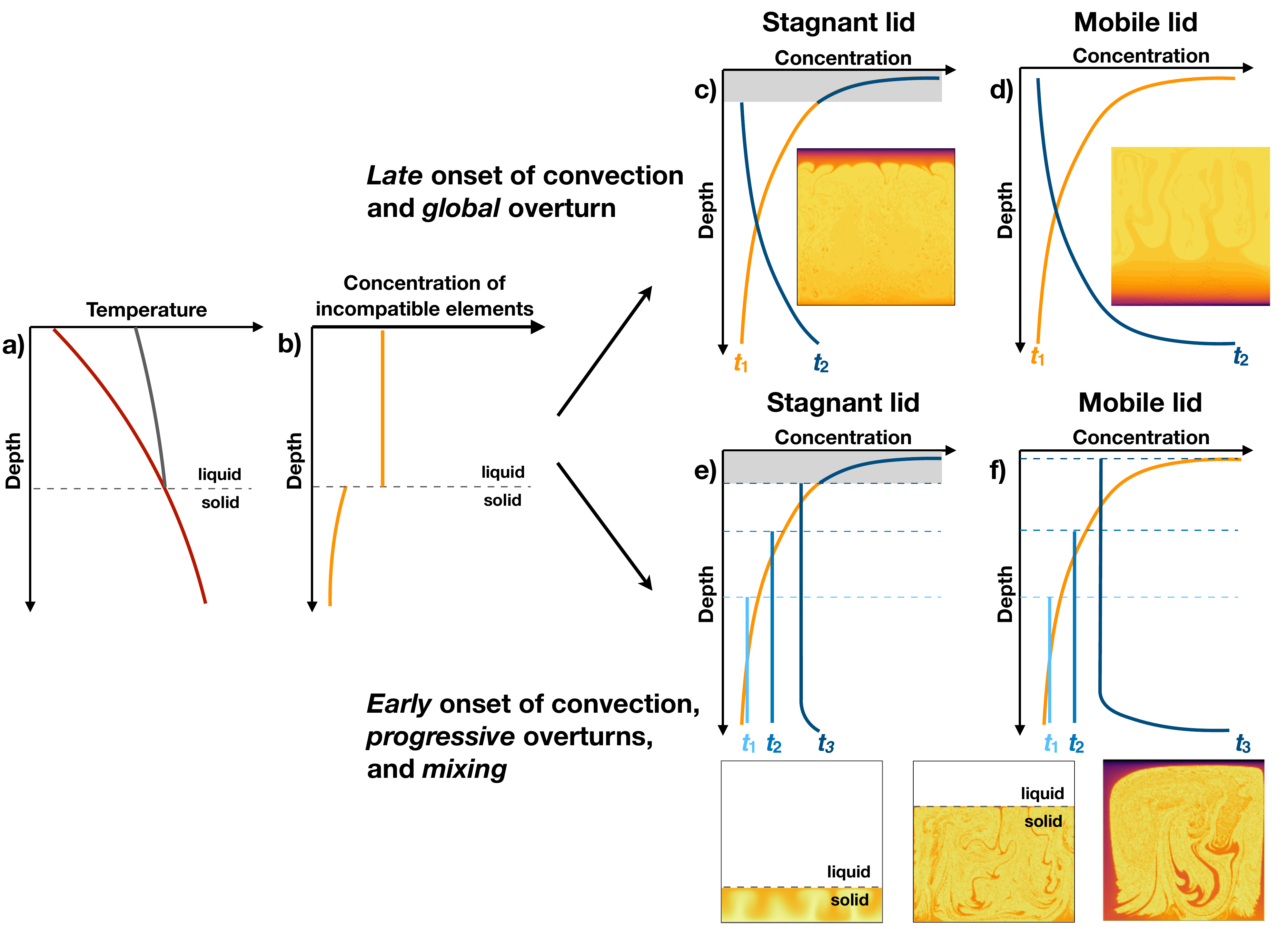}
\caption{\add{Different scenarios for the consequences of the fractional crystallization of a magma ocean. a) Bottom up solidification due to the steeper slope of the liquid adiabat (grey line) with respect the melting temperature (red line). b) Enrichment of incompatible elements in the liquid phase. c), d) Complete fractional crystallization before the  onset of  mantle convection. \addd{FeO} and HPEs are more and more enriched at shallow depths (orange lines at time $t_1$). The formation of a stagnant lid (grey area in panel c) can prevent late-stage cumulates from sinking. The overturn takes place beneath the lid (blue line at time $t_2$ and contour plot in panel c). If the uppermost cumulates are mobilized, a global-scale overturn takes place  (blue line at time $t_2$ and contour plot in panel d). The contour plots schematically describe the concentration of \addd{FeO} and HPEs in the solid mantle, with low and high values in light yellow and black, respectively. e), f) As in panels c and d, but for the case of solid-state convection beginning during solidification. Times $t_1$ to $t_3$ refer to subsequent stages. An early onset of convection can cause mixing of newly-formed cumulates beneath the magma ocean. The formation of a stagnant lid leaves a largely mixed mantle depleted in incompatible elements (blue line at time $t_3$ in panel e). Surface mobilization causes the formation of a dense primordial layer, overlaid by a compositionally-homogeneous mantle (blue line at time $t_3$ in panel f). Snapshots in panels c and d are derived from simulations similar to those of \citet{Tosi2013b}; snapshots beneath panels e and f are obtained following the approach of \citet{Maurice2017}.}}
\label{fig:overturn}
\end{figure}

Because of the low viscosity of silicate liquids \citep[e.g.,][]{Karki2010}, turbulent convection initially controls the dynamics of magma oceans, which are expected to be well mixed and adiabatic \citep[e.g.,][]{Solomatov2015}. For relatively small bodies like Mercury, the Moon and Mars, the mantle melting temperature increases with pressure more rapidly than the convective adiabat. Upon cooling, solidification proceeds thus from the base of the magma ocean upwards (Figure \ref{fig:overturn}a). Given the shape of the relevant high-pressure melting curves \citep{Fiquet2010,Andrault2011}, this scenario likely applies also to the Earth, although the possibility exists for solidification starting at mid-mantle depths \citep{Stixrude2009}, ultimately leading to the formation of a basal magma ocean \citep[see ][for a review]{Labrosse2015}. For simplicity, here we do not distinguish between solidus and liquidus and simply consider the mantle melting temperature as the temperature corresponding to the so-called rheologically critical melt fraction. Above this threshold, which is usually set around  30--40\% \citep[e.g.,][]{Costa2009}, 
the crystal-melt mixture exhibits a liquid-like behavior; below it, crystals tend to form a stress-supporting interconnected network with partially molten rocks effectively deforming as a solid \citep[see e.g.,][]{Solomatov1993a}.

Depending on whether newly-formed crystals settle or are suspended by the turbulent flow, the magma ocean can undergo fractional or equilibrium (batch) crystallization. In the first case, liquid and crystals effectively separate without equilibrating. The composition of the liquid evolves continuously through the removal of solidified materials, ultimately leading to the silicate differentiation of the solid mantle (Figure \ref{fig:overturn}b). In the second case, crystals and liquid remain in equilibrium resulting in a compositionally-homogeneous mantle. The conditions determining whether settling or entrainment occurs are still poorly understood. The density difference between liquid and solids, the size of crystals, the vigour of convection, \add{the effects of planetary rotation,} and the cooling rate of the magma ocean, all of which evolve in complex ways during solidification, may play a role in controlling whether fractional or equilibrium crystallization occurs \citep[e.g.,][]{Tonks1990, Solomatov1993a, Solomatov1993b, Suckale2012, Solomatov2015, Maas2015, Cassanelli2016, Maas2019}. 

The Moon represents the paradigm of terrestrial body that likely experienced a global-scale magma ocean that underwent fractional crystallization \citep[e.g.,][]{Warren1985}. 
The analysis of Apollo samples and lunar meteorites, as well as remote-sensing measurements revealed that the ancient lunar highlands largely consist of anorthosites composed of plagioclase feldspar,
building a 30--40 km thick crust \add{as inferred from the inversion of gravity and topography data from the GRAIL mission} \citep{Wieczorek2013}.
In a lunar magma ocean that fractionally solidifies from the bottom up, plagioclase begins to crystallize at a pressure of around 1 GPa, when 70--80\% of the magma ocean is solid \citep[e.g.,][]{Snyder1992,Elkins-Tanton2011,Charlier2018}. At these conditions, the residual liquid is denser than plagioclase crystals. These float to the top of the magma ocean forming the bulk of the Moon’s anorthositic crust, \add{which dramatically reduces the heat loss at the surface and retards magma ocean solidification \citeg{Elkins-Tanton2011}} (see also Sections \ref{Sec__EarlyOnset} and \ref{Sec__PrimaryCrust}). 
Plagioclase flotation is thought to work efficiently only on the Moon or smaller bodies; for larger bodies like Mars, the more rapid increase of pressure with depth causes plagioclase to become stable at shallow depths where the magma ocean has already reached a degree of crystallinity high enough to prevent flotation \citep{Brown2009,ElkinsTanton12}. 

Mercury’s surface is unusually dark \citep[e.g.,][]{Robinson2008}. 
Spectral measurements \citep{Murchie2015,Peplowski2016} 
indicate that graphite is the material responsible for this characteristic. Indeed, \citet{VanderKaaden2015} showed that for Mercury’s melts, graphite is the only mineral that could form a flotation crust in a crystallizing magma ocean. 
Yet, the composition of Mercury’s surface is largely the product of secondary volcanism (Section \ref{Sec__SecondaryCrust}). Widespread lava flows buried over time the primary graphite crust, which seems to have been exposed in some craters. 
Similar to Mercury, the surface of Mars is also largely volcanic but lacks evidence for primordial materials generated through the fractional crystallization of a magma ocean (Section \ref{Sec__SecondaryCrust}).

From the perspective of the convective dynamics and evolution of the solid mantle, the fractional crystallization of a magma ocean can have fundamental consequences beyond the formation of a primordial crust. As solids are removed, incompatible elements tend to be progressively enriched in the remaining liquid phase. These include \addd{w\"ustite (FeO)}
 and long-lived heat-producing elements (HPEs) Uranium (U), Thorium (Th) and potassium (K) in addition to volatiles such as water (H$_2$O) and carbon dioxide (CO$_2$). 
The concentration $C_\text{sol}$ of incompatible elements at a given radius $r$ of the solidified cumulates can be expressed as follows \citep[e.g.,][]{Boukare2018,Morison2019}:
\begin{equation}
    C_\text{sol}(r) = DC^0_\text{liq}\left( \frac{R_\text{p}^3 - R_\text{b}^3 }{R_\text{p}^3 - r^3} \right)^{1-D},
\end{equation}
where $C^0_\text{liq}$ is the initial concentration in the liquid magma ocean, $R_\text{p}$ and $R_\text{b}$ are the top and bottom radii of the magma ocean (corresponding to the planet radius and, in the case of a global magma ocean, to the core radius), and $D$ is a partition coefficient, typically $\ll 1$. The concentration of incompatible elements in solidified cumulates tends thus to follow a power law profile (Figure \ref{fig:overturn}b).

\subsection{Late onset of mantle convection and global overturn} \label{Sec__LateOnset}

If fractional crystallization proceeds fast compared to the timescale for the onset of solid-state convection (see below), the progressive enrichment of \addd{FeO} in the residual liquid ultimately causes the formation of dense, \addd{FeO}-rich layers at shallow depths overlying deep, \addd{FeO}-poor layers \citep{Elkins-Tanton2003,Elkins-Tanton2005b,Tosi2013b,Plesa2014,Scheinberg2014}. This configuration is gravitationally unstable and prone to overturn (Figures \ref{fig:overturn}c and \ref{fig:overturn}d).
The scenario of fractional crystallization followed by a global-scale overturn has been proposed for Mercury \citep{Brown2009}, Mars \citep{Elkins-Tanton2003,Elkins-Tanton2005b,Elkins-Tanton2005c}, the Earth \citep{Elkins-Tanton2008,Foley2014}, and the Moon, although in the latter case, the overturn  only involves a dense layer forming beneath the plagioclase crust \citep[][]{Zhong2000,Zhang2017,Li2019,Yu2019,Zhao2019} (see also Section \ref{Sec__Crust}).  

Flotation crusts or solid cumulates that crystallize at shallow depth and relatively low temperature have a high viscosity that prevents deformation. Late-stage cumulates enriched in \addd{FeO} and HPEs tend to remain locked in the stagnant lid with the overturn involving only the sub-lithospheric mantle (Figure \ref{fig:overturn}c). A stagnant lid with primordial composition thus forms that overlies a mantle with a moderately stable compositional stratification. Alternatively, the viscosity of the near-surface mantle may remain relatively low due the presence of interstitial melt or to a high surface temperature, allowing for sluggish mobilization \citep[e.g.,][]{Scheinberg2014}. Alternatively, the stresses induced by the first upwellings impinging at the base of the lithosphere may locally exceed its yield strength, inducing an early episode of plate-tectonics-like surface mobilization \citep{Tosi2013b}. In both cases a whole-mantle overturn takes place, resulting in a highly stable compositional gradient due to increasing \addd{FeO}-enrichment of the overturned materials from the surface to the base of the mantle (Figure \ref{fig:overturn}c). 


The global-scale overturn of a fractionally crystallized magma ocean---beneath the stagnant lid or involving the lid itself---has been studied in particular in the framework of the early evolution of Mars as it provides a suitable explanation for the generation of an early magnetic field \citep{Elkins-Tanton2005c,Plesa2014,Scheinberg2014} (see Section \ref{Sec__MagField}), for the rapid formation of secondary crust caused by upwelling cumulates undergoing partial melting upon overturning \citep{Elkins-Tanton2005b} (see Section \ref{Sec__Crust}), and for the generation of compositionally-distinct domains in the silicate mantle \citep[e.g.,][]{Elkins-Tanton2003, Debaille2009}. The latter constraint is derived from the analysis of Martian meteorites whose isotopic heterogeneity requires the formation of distinct source-reservoirs within the first $\sim$100 Myr of the solar system that remained unmixed throughout the planet's evolution \citeg{Mezger2013}.

Albeit successful, a global-scale overturn of the above kind 
poses a problem for the long-term evolution of the Martian mantle. As discussed in Section \ref{Sec__Crust}, surface evidence indicates that Mars has been volcanically active throughout its history \citep{Hartmann1999,Werner2009}. Long-lived volcanism in a stagnant lid body like Mars is only possible in the presence of a convective mantle where hot plumes rising from the core-mantle boundary (CMB) undergo decompression melting \citep{Plesa2018}. Using the crystallization sequence proposed by \citet{Elkins-Tanton2005b}, \citet{Plesa2014} performed numerical simulations of global-scale overturn in the Martian mantle in the presence of either a stagnant or a mobile lid. They showed that in both cases the compositional gradient following the overturn (Figure \ref{fig:overturn}c and \ref{fig:overturn}d) is stable enough to largely prevent the subsequent onset of thermal convection and partial melting, at odds with the evidence that volcanism lasted over most of Mars' history, until as recently as few million years ago \citep{Neukum2004}.

The reason for the above behaviour is to be sought in the large buoyancy ratio associated with the overturned mantle. The buoyancy ratio ($B$) is defined as
\begin{equation}
    B =\frac{\Delta\rho_C}{\Delta\rho_T}=\frac{\Delta\rho_C}{\rho_0\alpha\Delta T}, \label{eq:B}
\end{equation}
where $\Delta\rho_\text{C}$ is the compositional density difference between shallow, \addd{FeO}-poor cumulates and deep, \addd{FeO}-rich cumulates that hinders convection, and $\Delta\rho_T$ is the density difference due to thermal expansion between cold shallow cumulates and deep hot cumulates that drives convection ($\rho_0$ is a reference density, $\alpha$ the coefficient of thermal expansion, and $\Delta T$ the temperature difference between the surface and the base of the mantle). \add{For a mantle that underwent a bottom-up solidification, $\Delta\rho_T$ in eq. \eqref{eq:B} can be well determined assuming that the initial core temperature corresponds to the melting temperature of the silicate mantle at CMB conditions. Much more uncertain is the value of $\Delta\rho_\text{C}$, which can range from 0 for the (unlikely) case of a complete equilibrium crystallization up to several hundreds of kg/m$^3$ in the case of a purely fractional crystallization.} The crystallization sequence obtained by \citet{Elkins-Tanton2005b} \add{assuming pure fractionation} and used by \citet{Plesa2014} results in a buoyancy ratio of $\sim$3, sufficiently large to suppress thermal buoyancy and solid-state convection in the overturned mantle. As shown by \citet{Tosi2013b}, post-overturn thermo-chemical convection begins to be possible (at least in the upper mantle) for values of $B$ lower than $\sim$2. 

In addition, the dense, \addd{FeO}-rich cumulates that sink to the CMB are also highly enriched in HPEs. In the presence of a high value of $B$ preventing the deep mantle from convecting, such cumulates will tend to heat up. The accompanying temperatures can then exceed the solidus, leading to the formation of partial melt, or even the liquidus, leading to the formation of a basal magma ocean \citep{Plesa2014,Scheinberg2018}, whose consequences for Mars are yet to be explored. 


\subsection{Early onset of mantle convection and progressive mixing} \label{Sec__EarlyOnset}

Convective mixing acting while the magma ocean is still solidifying provides a viable way to reduce the effective buoyancy ratio of a solid mantle formed via fractional crystallization. The scenarios described above apply whenever the timescale of magma ocean solidification is shorter than the timescale of convective mobilization of newly formed cumulates. The presence of a growing atmosphere on top of a crystallising magma ocean has been widely recognized as capable to exert a fundamental control on the magma ocean lifetime \citeg{Abe1986,Elkins-Tanton2008,Lebrun2013,Hamano2013,Nikolaou2019}. Similar to \addd{FeO} and HPEs, volatiles are enriched in the liquid phase as the magma ocean cools and solidifies, with their concentration that can increase until reaching saturation in the magma. In this context, H$_2$O and CO$_2$ are the two gases that have been studied most extensively. The amount of gas in excess of saturation is generally assumed to form bubbles that efficiently rise to the surface and escape from the magma ocean forming a secondary atmosphere. This slows down the solidification due to the ability of H$_2$O and CO$_2$ to act as greenhouse gases absorbing infrared radiation. The solidification timescale depends on several poorly constrained parameters \citep{Nikolaou2019}. Above all, it is influenced by the initial bulk abundance of volatiles and by the incoming stellar radiation \citep[e.g.,][]{Hamano2013,Lebrun2013}. For the Earth, for example, model calculations show that a volatile-depleted, whole-mantle magma ocean would solidify in only $\sim$10$^3$ years \citep[e.g.,][]{Monteux2016}. By contrast, the presence of an outgassed steam atmosphere corresponding to a bulk water inventory of one Earth’s ocean would extend the solidification to more than 1 Myr \citep[e.g.,][]{Elkins-Tanton2008,Lebrun2013,Salvador2017,Nikolaou2019}. 
Besides the presence of a thick outgassed atmosphere, the formation of a solid flotation crust can slow down magma ocean solidification even more significantly. In the case of the Moon, a growing flotation crust forming a solid conductive lid on top of the crystallizing magma ocean extends its solidification timescale to at least few tens of millions of years \citep{Meyer2010,Elkins-Tanton2011,Perera2018}. For Mercury, the idea of a graphite flotation crust is relatively new, the thickness of such crust highly uncertain \citep{VanderKaaden2015}, and dynamic models of Mercury's convection accounting for the effects of magma ocean solidification still absent. Therefore it remains unclear what its effect would be. 

At any rate, a slowly solidifying magma ocean opens the possibility for convection to set in during solidification. Newly formed solid cumulates can overturn and start to be mixed by convection while being overlaid by a liquid magma ocean (Figure \ref{fig:overturn}e and \ref{fig:overturn}f). If mixing is efficient, two end-member configurations at the end of solidification are possible, similar to the previous case of a global overturn. The latest, highly enriched cumulates can remain locked in the stagnant lid, while the underlying mantle has been largely homogenized by an early onset of convection (Figure \ref{fig:overturn}e). Alternatively, if the lid can be mobilized, the uppermost cumulates can sink to the CMB forming a basal layer of materials enriched in incompatible elements (Figure \ref{fig:overturn}f).

\begin{figure}[ht!]
\centering
\includegraphics[width=0.7\textwidth]{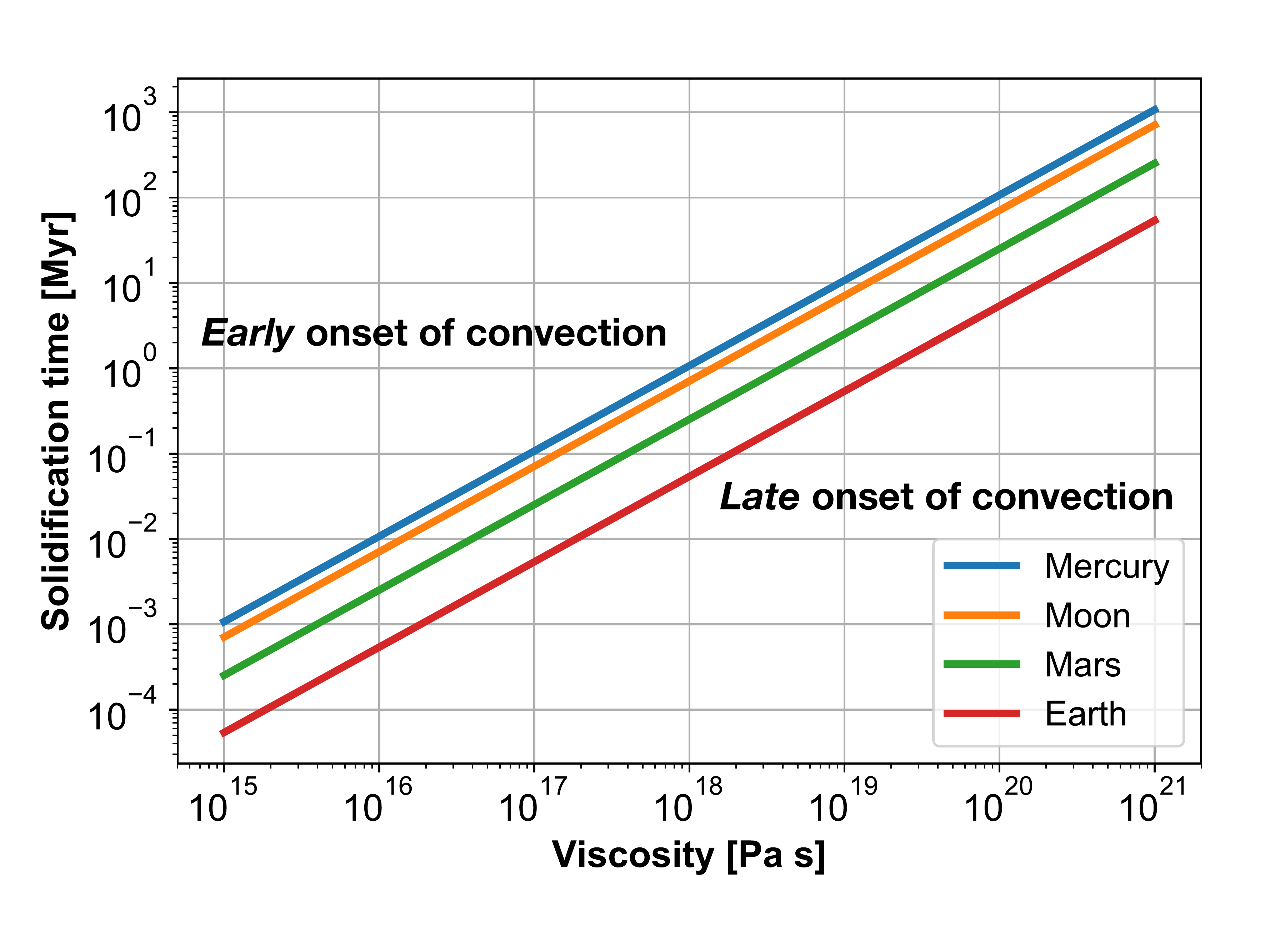}
\caption{Boundary separating the regime of early onset of convection (above the curve) and late onset of convection (below the curve) for different bodies according to magma ocean solidification time and viscosity of the solid mantle. For a given value of the viscosity, the solidfication time allowing for an early onset of convection is longest for Mercury and shortest for the Earth. Similarly, for a given solidification time, the viscosity making an early onset possible is lowest for Mercury and largest for the Earth. The diagram is based on eq. \eqref{eq:Rc} and calculated assuming $R_C=5\times 10^4$.}
\label{fig:timescale}
\end{figure}

Whether or not the solid mantle begins overturning before the end of magma ocean crystallization largely depends on a competition between the timescale of solidification \add{($\tau_\text{MO}$)} and the timescale of convective destabilization of the mantle \add{($\tau_\text{conv}$)} \add{due to the unstable compositional stratification} \citep{Maurice2017,Ballmer2017,Boukare2018}. To quantify the relation between the two timescales, \citet{Boukare2018} introduced a non-dimensional number ($R_C$) defined as
\begin{equation}
    R_C = \frac{\tau_\text{MO}}{\tau_\text{conv}} =  \tau_\text{MO} \frac{\Delta\rho_C g D}{\eta}, \label{eq:Rc}
\end{equation}
where $\tau_\text{MO}$ is the magma ocean solidification time, $\eta$ the dynamic viscosity of the solid mantle, $\Delta\rho_C$ the density difference between \addd{FeO}-rich and \addd{FeO}-poor cumulates (as in eq. \eqref{eq:B}), $g$ the gravity acceleration, and $D$ the thickness of the solidified mantle. The larger $R_C$ is, the more likely is that convection will start mixing the mantle during its solidification. As shown by \citet{Boukare2018} in the framework of isoviscous models, the critical value of $R_C$ that separates the two scenarios of late onset of convection and global overturn (Figures \ref{fig:overturn}c and \ref{fig:overturn}d) as opposed to an early onset characterized by progressive overturns (Figures \ref{fig:overturn}e and \ref{fig:overturn}f) is between $10^4$ and $10^5$. Figure \ref{fig:timescale} (which reproduces Figure 3b of \citet{Boukare2018}), shows the boundary separating the two regimes assuming $R_C=5\times 10^4$, $\Delta\rho=1000$ kg/m$^3$, and for $g$ and $H$ the gravity acceleration and mantle thickness of the corresponding body. As shown by eq. \eqref{eq:Rc}, an early onset is facilitated by long solidification times, a thick mantle and a low viscosity. 

The presence of residual melts as well as volatiles renders low values of the viscosity ($10^{17}$ Pa s or even lower) fully plausible \citep[see, e.g.,][]{Maurice2017}. For these viscosities, an early onset of convection and mixing (Figures \ref{fig:overturn}e and \ref{fig:overturn}f) becomes likely for Mars \citep{Maurice2017} in the presence of an outgassed atmosphere retarding magma ocean solidification \citeg{Lebrun2013}---a scenario that \citet{Ballmer2017} has proposed also for the Earth---and inevitable for the Moon \citep{Boukare2018} because of the strong insulating effect of the flotation crust, which remained part of the stagnant lid. \add{Furthermore, recent work that considers the effects of a semi-permeable boundary at the solid-liquid interface via a melting-freezing boundary condition \citeg{Labrosse2018,Morison2019} suggests that the critical Rayleigh number for the onset of solid-state mantle convection can be significantly lower and the heat transfer efficiency significantly higher than for non-penetrative boundaries as assumed in the works of \citet{Maurice2017}, \citet{Ballmer2017} and \citet{Boukare2018}. In such a case, an early onset of convection would be facilitated and its effects strengthened.} Although a flotation crust was probably present also on Mercury (see also Section \ref{Sec__Crust}), the low convective vigor due to the planet's thin mantle makes the hypothesis of an early onset of convection unlikely and the possibility of a global overturn beneath the stagnant lid and flotation crust \citep{Brown2009} more plausible (Figure \ref{fig:overturn}c). The inferred lack of a highly enriched primary crust on Mars (Section \ref{Sec__Crust}) suggests that an early episode of surface mobilization accompanied by the subduction of the latest cumulates (Figure \ref{fig:overturn}f) may have characterized the final stages of magma ocean solidification. 

\section{Crustal manifestations}\label{Sec__Crust}
%
Being their geologic evolution largely the result
of interior processes, planetary crusts contain important clues
on the nature and temporal variations
of such processes.
The focus here is on three aspects of the 
crust: type, volume, and age of emplacement.


%


Typically, three types of crust
are identified 
according to a temporal criterion, which
also discriminates the different processes
associated with their production \citeg{Taylor2009}.
As anticipated in Section \ref{Sec__MagmaOceans}, primary crusts are the result of magma
ocean solidification and are emplaced
very early in the history of an object;
secondary crusts are the result of 
partial melting of the mantle, which can occur
throughout the solar system history;
and tertiary crusts are the results of, e.g.,
remelting of the secondary crust, and at
present \add{the only} concrete 
evidence for tertiary
crust is 
\add{the continental crust of the} Earth, possibly
because plate tectonics is required to
produce it 
\citep{Taylor2009}.

Characterizing the type of crust(s) present
in a given body has important implications
for models of its thermal evolution, 
since the crust 
is chemically different from the bulk
composition, most importantly in its
content of the long-lived HPEs.
Thus, crustal production effectively
decreases the value of $H,$ the specific
(i.e., volumetric) rate of internal heat generation appearing
in the Rayleigh number for internally heated
convection \add{\citeg{Turcotte2002}}:
\begin{equation}
    {\rm Ra}_{\rm H}=\frac{\rho g\alpha D^5H}{\eta k \kappa}.
\end{equation}
In this expression $\rho$ and $D$ are the density and
thickness of the mantle, $g$ the gravitational
acceleration; $\alpha$, $\eta,$ and $\kappa$
are the thermal expansivity, \add{conductivity}, 
and diffusivity; $\eta$ is the dynamic 
viscosity.
In the case of an early emplaced primary crust, 
the effect on the internal evolution 
can be treated as an
initial boundary condition
on the value of $H$
\citeg{Laneuville2013,Rolf2016}.
In the case of a secondary crust,
which in general is created over an
extended period of time,
the value of $H$ has to be time-dependent,
consistent with the secondary
crust produced \citeg{Morschhauser2011,Padovan2017}.
The absolute amount of heat sources
that are extracted in creating crustal
material depends critically on the
volume of this material and
its enrichment with respect to the
mantle.
These two parameters should be
characterized for each crustal
type and, at least for the secondary
crust, which is emplaced over an
extended period of time, for its 
different units.

The bulk volume of the crust can 
be investigated using a combination
of gravity and topography data.
The basic idea is that the non-central 
components of the gravitational
field of a planet contain the contribution
of the mass associated with the topography and
with any internal non-radially symmetric
mass distribution.
By ascribing the Bouguer anomaly---the
gravitational signal left after the 
removal of the topographical 
signal---to subsurface
interfaces or loading, models for the 
thickness of the crust can be obtained.
This is usually done either by assuming
a given compensation mechanism
\citeg{Wieczorek97,Padovan2015}, or by
globally mininizing the anomalies 
using an anchor point obtained, 
e.g., by imposing an
average thickness of the crust or
by requiring the thickness
to be larger than zero everywhere
\citep[see, e.g., the review by][]{Wieczorek2015}.
In addition to the bulk volume, it is
important to identify the volumes of different
units that are associated with different
crustal types and/or different time of
emplacement.
As a case in point, the lunar maria are 
dark basaltic provinces located mostly 
on the near-side of the Moon.
They represent the extruded component 
of magmatic events
associated with secondary crustal production,
and thus HPEs mantle depletion.
In principle, surficial units can be dated 
through crater counting (Section \ref{Sec__Impacts})
and their volumes can be estimated by  
taking advantage of additional stratigraphical, 
compositional, and geological analyses
\citeg{Ernst2015,Whitten2011}.

The enrichment is typically defined with respect
to some model for the bulk abundance of 
heat-producing elements. 
The availability
of lunar and martian meteorites
greatly improves
estimates of the bulk abundances
of these bodies 
\citeg{Taylor2013,Taylor2014}.
In the case of Mercury the 
bulk composition of the silicate
part is obtained from a combination 
of geochemical and geophysical considerations
informed by experiments, 
analysis of the surface composition, and formation 
and early evolution scenarios
\citep[see][for a recent review]{Nittler2018}.
The enrichment of the primary crust  
depends on a number of physical parameters
both extensive---e.g., pressure profile of the body
and bulk composition of the magma ocean---and 
intensive---e.g., density and composition
of the last cumulates to solidify.
For the secondary crust, 
which results from the 
accumulation of melt produced
at different times and different
locations in the mantle,
the enrichment is better
defined as a function of local thermodynamic
conditions (pressure, temperature, composition,
melt fraction), which in turn are a function
of the overall evolution of the body.

\subsection{Primary crusts}
\label{Sec__PrimaryCrust}

Primary crusts are formed as a result of the late-stage
solidification of a magma ocean 
(Section \ref{Sec__MagmaOceans}).
While it is expected that bodies at least as
large as Vesta 
\add{(i.e., with an equivalent diameter larger
than about 500 km)} go through a magma ocean
phase \citep{ElkinsTanton12}, thus
possibly producing a primary crust, 
for bodies larger than Mercury this 
primary crust is likely overturned and
thus, it is not directly observable
at present time.
The best known, 
and possibly only certain, example of a 
primary crust
is represented by 
the lunar anorthositic crust \citep{Wood1970}, 
which is enriched in HPEs \citep{Taylor2014}.
In a sense, the lunar primary crust is also
an exception, since it represents a flotation
crust, created by the upward accumulation 
of light crystals during the 
last phases of solidification of
the lunar magma ocean.
It is generally accepted that also Mercury
underwent an initial hot phase characterized
by the presence of a global magma
ocean \citeg{Brown2009}.
In the case of Mercury the only mineral
that could float 
and thus originate a lunar-like,
gravitationally stable
primordial crust is graphite, a very
dark mineral
\citep{VanderKaaden2015}.
Overall, the crust of Mercury as observed
today is interpreted as being 
for the most part secondary
\citeg{Denevi13}, with its 
oldest units having been emplaced in 
a period consistent with the timing of the
late heavy bombardment (LHB) \citep{Marchi2013}.
These findings indicate that any
primary crust would be difficult to
identify.
However, the general dark appearance of 
the Mercurian surface and the
existence of extensive dark units, 
the so called low reflectance material
\citeg{Murchie2015}, may
represent the sign of a dark, 
graphitic primary crust subsequently
covered by volcanic material and
in turn reworked by the intense 
period of bombardments typical of
LHB scenarios \citep{Ernst2015,Peplowski2016}.
The possible volume of the primary crust 
depends on the unknown bulk carbon content
of Mercury.
If the carbon content of Mercury lies anywhere
between the carbon-poor silicate portion of the Moon
and the carbon-rich CI meteorites, then
the thickness of the Mercurian primary crust
can be anywhere between 1 cm and 10 km
\citep{VanderKaaden2015}.

The lunar anorthositic crust and the putative
graphite primary crust of Mercury are 
easy to label as primary crusts, since
they result from floatating, 
chemically well-defined minerals.
However, the definition of what 
constitutes a primary crust
is both wider and not necessarily 
unique.
What is commonly accepted is that
an initial magma ocean phase characterizes
all planetary-sized bodies, and 
the evolving composition of these 
solidifying magma oceans results 
in \addd{FeO}- and HPE-rich late cumulates. 
Under the simplifying assumption
that no convection operates in the solid 
mantle while the magma ocean is still 
present (but see Section \ref{Sec__EarlyOnset}),
the resulting radial density profile in
the solid mantle is 
gravitationally unstable
\citeg{Brown2009,Elkins-Tanton2011}.
Apart from the light components, as in the 
case of the Moon and probably Mercury,
this primary heavy crust is likely to be
overturned, i.e., to sink at larger depths
until a gravitationally stable profile
is reached.
Indeed, crystallization models for the lunar
magma ocean predict the formation of a dense
ilmenite-rich layer \citeg{Elkins-Tanton2011}, 
which is overlain by the floatation crust 
(or explicitly, by the ``floatating component'' of
the primary crust).
The chemical signature of this layer would
make it part of the primary crust.
However, its high density and possibly low
viscosity would make it sink in the mantle
possibly reaching all the way to the core-mantle
boundary \citeg{Parmentier2002,Yu2019}.
Depending on its geometry, this overturn
could explain the asymmetric nature of the
lunar secondary crust (Section \ref{Sec__SecondaryCrust}
below).

The rapid accretion of Mars indicates that
it sustained a global magma ocean
\citep{Dauphas2011}.
Models for the radial density profiles
resulting from the solidification of a
Martian magma ocean  
do not predict the formation of
a floatation crust and
show a strongly gravitationally unstable 
profile, which would likely 
rapidly overturn through a 
Rayleigh-Taylor instability
\citep{Elkins-Tanton2003}.
As shown in Section \ref{Sec__MagmaOceans}, since these heavy components of the 
primary crusts are rich in HPE, their overturn 
may segregate part of the heat-sources at the bottom
of the mantle, as also proposed---but as of yet
\add{lacking observational
verification}---for the Moon,
the Earth, and Mars
\citeg{Labrosse2007,Zhang2013,Plesa2014,Yu2019,Li2019}.

\subsection{Volume and time of emplacement of the 
secondary crust}\label{Sec__SecondaryCrust}

Contrary to the case of the 
Earth, where plate tectonics
operates, melting activity
in one-plate bodies is associated
only with decompression melting in
mantle upwellings
\citeg{Baratoux2013}.
The melt, if buoyant, contributes
to the formation and thickening of the
so-called secondary crust
\add{through eruptions and/or intrusions}.
Thus, the volume of the secondary crust
as observed today represents the 
cumulative amount of melt produced
in the mantle 
during the 4.5 Gyr of evolution of 
the solar system.
The thermal state of the mantle
largely controls the secondary
crustal production 
\citeg{Laneuville2013,Grott2013,Padovan2017},
so putting constraints on  
the temporal evolution of secondary 
crustal building
would indirectly provide insight
on the thermal evolution of the 
mantle.
This approach rests on the 
possibility of dating the various 
surface units, typically through
crater counting 
(Section \ref{Sec__Impacts}),
and  of interpreting them 
in terms of interior processes.

The surface of the Earth is 
continuously reworked by plate tectonics
and erosion, and the dating of geological
event gets more difficult with older ages,
in particular in the Precambrian era, 
which ended about 600 Ma
\citeg{Press1978}.
Similarly, Venus has a surface younger
than about a billion years \citeg{Hauck1998}, 
which is interpreted as being 
the result of episodic catastrophic 
overturns \citeg{Rolf2018} or
continuous volcanic resurfacing
\citeg{King2018}.
It is therefore difficult to reconstruct
the ancient volcanic history of these two bodies.
On the contrary, the geological record 
of the Moon and Mercury, 
airless bodies with no plate tectonics,
potentially holds the signs 
of events that occured
during the entire solar system evolution.
Mars does not have plate tectonics, 
unless perhaps for a short time early after
its formation \citep[e.g.,][]{Nimmo2000} (see also Sections \ref{Sec__MagmaOceans} and \ref{Sec__MagField}), but
early in its evolution
its surface has been altered by
aeolian, fluvial, glacial, 
and possibly lacustrine---i.e.,
exogenic---processes
\citeg{Craddock2002,Bibring2006}
and the interpretation of its 
geological record is significantly 
more complicate than for
Mercury and the Moon
\citeg{Carr2006}.

As described above, the
crust of the Moon is for 
the most part primary in 
nature.
The dark patches that can 
be observed with the naked-eye on the near
side of the full Moon represent 
the surficial manifestations of
magmatic events associated with 
the relatively minor 
secondary lunar crust production.
Overall they accounts 
for only about 0.7\%
of the volume of the crust 
\citeg{Head1992}. 
Typically, every surficial---or
extrusive---volcanic event is associated with 
a subsurface reservoir of magma, 
its intrusive counterpart.
The amount of intrusive volcanism is 
difficult if not impossible to assess
remotely, and on Earth, where
a lot of data are available, 
the variability of the intrusive-to-extrusive
ratio is large \citep{White2006}. 
Typically a value of 10 times the 
extruded component can be used
as a very rough rule of thumb.
Accordingly, the crust of the Moon has 
a secondary component corresponding to
about 7\% of its volume.
This secondary component, as inferred
from crater counting and analysis of
samples returned during the 
Apollo program, shows that the 
vast majority of the lunar 
secondary crust has been emplaced
in the first 1.5 billions of years
of evolution 
\citep[e.g., Fig. 18 of ][]{Hiesinger2011}.

The secondary crust of the Moon is peculiar
in its highly asymmetrical distribution.
Almost the entire majority of 
secondary crustal units are located
in the near side, mostly within
large impact basins
\citeg{Hiesinger2011}.
The location of the secondary crust
coincides with a region of the surface
rich in incompatible elements like
Th and U
\citeg{Lawrence1998,Lawrence2003},
which encompasses the Procellarum and
Imbrium basins and is commonly
referred to as Procellarum
\add{KREEP} Terrane (PKT), \add{because of its strong enrichment in potassium (K), rare-earth elements (REE) and phosphorous (P)} \citep{Jolliff2000}.
Additionally, recent magnetic data
obtained by the Lunar Prospector
and Kaguya missions \citep{Tsunakawa2015},
show a region of weak crustal 
magnetization roughly corresponding
to the PKT \citep{Wieczorek2018}.
\add{The high crustal temperatures associated
with a higher abundance of heat producing
elements in the PKT region could have 
prevented retention of a magnetic signature
\citep{Wieczorek2018}.}
The above set of observations 
have been interpreted
as indicating the presence of 
material enriched in HPE
in the crust or below the crust
in the PKT region
\citep{Laneuville2013,
Laneuville2018}.
In addition to being compatible
with the measurements listed above,
this scenario can explain
a number of additional observations, 
including asymmetric lunar impact basin 
morphologies \citep{Miljkovic13} and
heat flux variations as directly measured at the
Apollo 15 and 17 landing sites and as remotely
inferred at one location 
using the Lunar radiometer experiment 
onboard the Lunar Reconnaissance Orbiter
\citep{Langseth1976,Warren1987,Paige2016}.
While a fully satisfactory explanation 
for the creation of an 
enriched layer localized in the
near side is currently lacking,
\add{several mechanisms have been
proposed, both endogenous and
exogenous.}

Volcanism that is localized on 
a single lunar hemisphere may be 
indicative of 
\add{an early phase of} degree-1
convection in the Moon, i.e., convection 
where buoyant material rises in a single
upwelling \citep{Zhong2000}.
This scenario may arise if the late 
dense cumulates 
resulting from the solidification of the 
lunar magma ocean and forming 
beneath the flotation crust 
would sink to the
CMB. 
For some combination of 
the rheological properties of the mantle,
this material 
would induced a single mantle upwelling
\citeg{Zhang2013}. 
In addition, part of these
heavy cumulates would remain at the
CMB up to the present day
\citep{Zhang2017,Zhao2019}, potentially
providing a mechanism to explain the
presence of partial melt there, as 
inferred from the Apollo seismic data
\citep{Weber11}.
As initial conditions these models 
typically assume a post-overturn
scenario, where the heavy late 
cumulates resulting from the 
solidification of the lunar magma
ocean have already sunk to the CMB.
However, as described in 
Section \ref{Sec__EarlyOnset}, the
mechanism of the overturn is  
debated and accordingly,
\add{its role in}
the asymmetrical nature of the lunar
evolution is an open
question.

\add{Alternatively, the dichotomy has
been ascribed to the effect of an impact.
\citet{Jutzi2011} investigated the slow-velocity 
accretion of a companion moon with
a mass about 4\% of the lunar mass and a 
diameter of 1270 km, formed
in the same protolunar disk and thus, with 
a composition similar to the Moon. 
Such a slow-velocity impact on 
the current far side would explain the 
thickening of the far-side lunar crust
and the displacement of a residual, KREEP-rich
magma ocean in the near side.
This scenario, while possible, 
would require the correct timing
of the impact with respect to the 
magma ocean solidification.
Some authors have speculated that
the compositional and topographical 
properties of the lunar near side
are compatible with an
ancient mega-basin 
\citep{Cadogan1974,Whitaker1981}.
Building on this hypothesis, 
\citet{Zhu2019} explained the observed
crustal dichotomy in thickness and
composition as the result of an impact
of an object with a diameter of about
750 km on the current near-side on an already
solidified Moon. 
The ejecta distribution would explain the
crustal thickness dichotomy, and since 
the excavated KREEP-rich cumulates located
below the pre-impact primordial crust would 
flow back in the resulting giant basin, this
model would provide an explanation for the 
exposure of KREEP-rich material in the
near side.
However, gravity gradiometry obtained from 
the GRAIL mission indicates that the
subsurface structure in the near side
is the result of endogenous magmatic-tectonic
structures, likely not compatible
with a large impact \citep{Andrews-Hanna2014}.
Furthermore, the composition of
the far side crust as predicted with the
giant impact scenario does not match the 
observed composition \citep{Zhu2019}.}

Similar to the Moon, the secondary crust 
of Mars has a striking asymmetry. 
There is a hemispheric difference of about 5 km 
in surface elevation between the ancient southern 
highlands and the younger northern lowlands.
This difference is recognized as one of the 
oldest features of Mars and is reflected in 
the so-called crustal dichotomy, where,
based on gravity and topography data,
the crust has an average thickness of 
about 45 km in the northern lowlands, but is 
about 25 km thicker under 
the southern hemisphere 
\citeg{Neumann2004,Wieczorek04,Plesa2016}. 
The origin of this difference is the 
subject of a vast literature. 
Several mechanisms have been investigated,
including exogenous processes involving a giant impact 
\citeg{Andrews-Hanna2008,Marinova2008} 
and endogenous ones associated with the 
formation---as proposed for the Moon---of a 
hemispherical (i.e., degree-1) upwelling 
\citeg{Roberts2006,Keller2009,Sramek2012}. 
Additional hypotheses include a hemispherical 
overturn of a crystallized magma ocean 
\citeg{Elkins-Tanton2003} as proposed
for the Moon \citeg{Parmentier2002}, 
or a combined exogenous-endogenous scenario,
with a giant impact triggering a degree-1 
convection planform \citep{Golabek2011,Citron2018}. 

Irrespective of the mechanism responsible 
for the generation of the dichotomy,
analysis of Martian meteorites indicates that the
crust of Mars is secondary, the bulk of which
has been emplaced very early on,
likely within 100 Myr of the planet formation 
\citeg{Nimmo2005}, and certainly no
later than the early Noachian
\citep[\addd{about 4.2 Ga,}][]{Grott2013}.
Based on the estimated volume of extruded
volcanic units \citep{Greeley1991}, 
and using a similar 
intrusive-to-extrusive ratio as for 
the Moon above, 
observations can account for about
7\% of the volume of the crust,
assuming its average thickness
is 62 km \citep{Wieczorek04,Plesa2018}.
While Mars shows signs of very recent
volcanic activity
\citep{Hartmann1999,Werner2009}, 
possibly extending
to the present day \citep{Neukum2004},
the bulk of the secondary crust
has been emplaced during the
first two billions of years of evolution
\citep{Greeley1991}, and mostly
very early on
as testified
by the geological record
in the Valles Marineris, which
shows that Mars underwent 
voluminous volcanism during
its first billion of years 
\citep{McEwen1999}.

The analysis of the data returned by the
spacecraft MESSENGER, which orbited
Mercury between March 2011 and
April 2015, indicates that
the surface of the innermost
planet likely represents the
prototype of a secondary crust,
since it is interpreted as 
being largely the result of volcanic 
events \citeg{Denevi13}.
By dating the different geological
units of the surface of the planet,
\citet{Marchi2013} showed that the 
oldest surfacial units have 
been emplaced between 4.0 and
4.1 \addd{Ga}.
This range of ages corresponds roughly with
the period of the LHB, and it has
been suggested that corresponding to
this time any record of older 
units would have been erased
due to the combined effect of 
impacts and possibly related 
volcanism \citep{Fassett12,Marchi2013}.
However, the observational 
framework supporting the LHB
scenario has been recently
put into question
\citep{Michael2018,Morbidelli2018}.
\add{Independent of the chosen
impactor flux, for which several
functional shapes have been 
proposed
\citeg{Neukum1994,Marchi2009,Morbidelli2012}, the oldest crust of
Mercury is younger than the oldest
Lunar crust.}
Given the secondary nature
of the crust, a straightforward 
explanation is that early on,
the hot post-formation mantle 
produced massive amounts of melt
and associated volcanic events
that simply kept obliterating 
older units.
However, through melt-induced
depletion of heat sources,
the mantle cooled
(see discussion on the 
time-dependent $H$ above)
and gradually melt-production decreased.
Using similar techniques to those
of \citet{Marchi2013},
\citet{Byrne2016} inferred that
the youngest large volcanic provinces
date back to 3.5 \addd{Ga}.
These timings are compatible with 
recent thermal evolution models 
of the planet 
\citep{Tosi2013,Padovan2017}.

\begin{figure}[t!]
\centering
\includegraphics[width=0.8\textwidth]{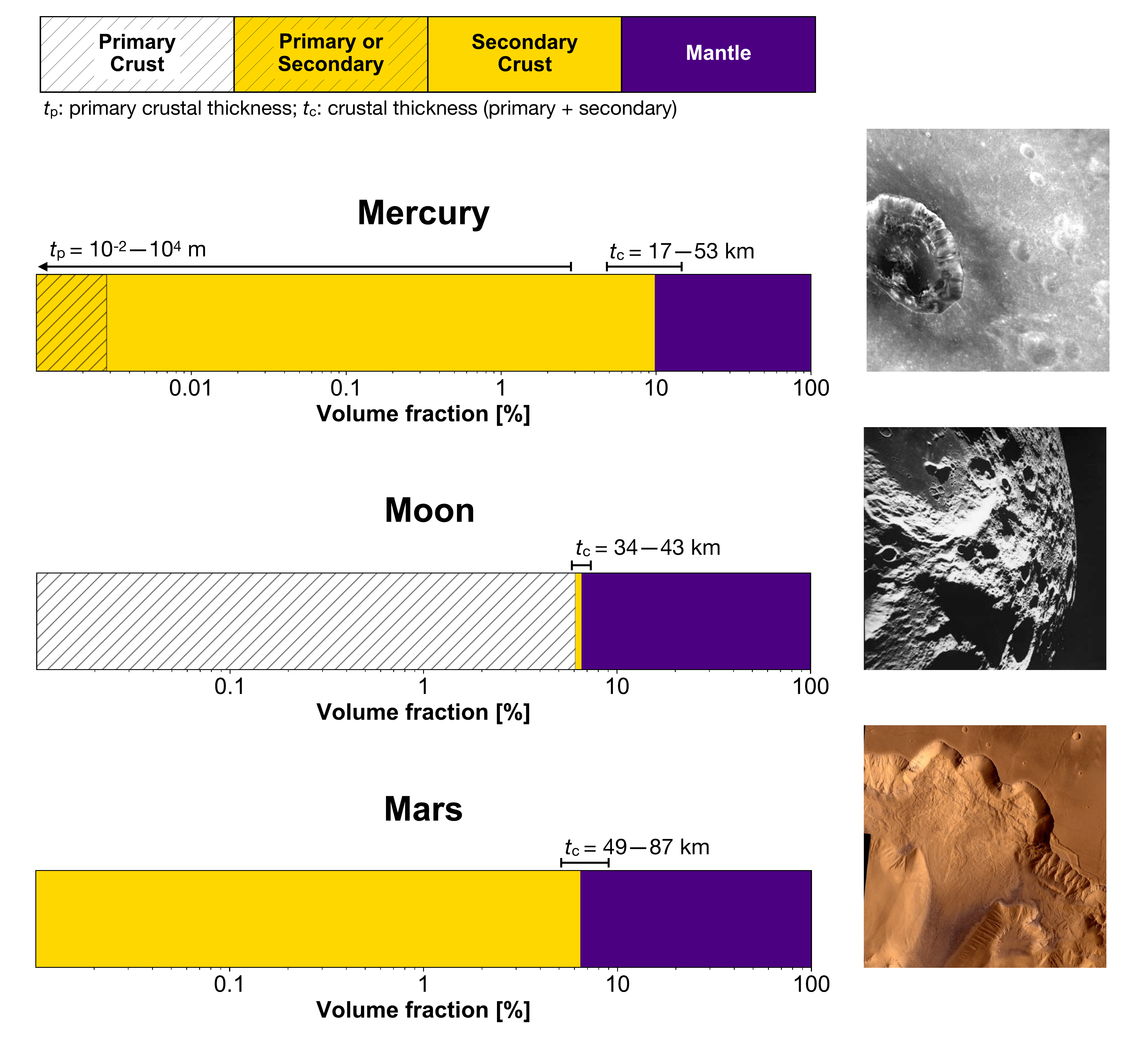}
\caption{Left column: Volumetric composition of the different silicate reservoirs of Mercury, the Moon, and Mars. Colors indicate composition, according to the legend on top. Note that the scale is logarithmic. The black bars indicate the range of values for the volume fractions, with the labels expressing the corresponding range of thicknesses. Sources for the data are described in the text.  Right column: example of the surfaces of each body. 
Mercury: Berkel crater (diameter $D=21$ km), image obtained by the MESSENGER mission. The background represents secondary crust, while the crater is a typical example of exposure of dark material, which has been ascribed to the planet's primary graphite crust. 
Moon: the crater on the top left is Aitken ($D=135$ km) on the far side of the Moon, in an image obtained by the Apollo 17 mission. The crater is infilled with dark volcanic material, i.e., secondary lunar crust. The light-coloured surrounding area is the primary anorthositic crust. 
Mars: Ophir Chasma, a valley connected to Valles Marineris, in an image obtained by the Viking 1 mission. This region is associated with massive volcanism early in the history of Mars. The large crater in the lower right corner is 30 km wide.
Image credits:
Mercury: NASA/Johns Hopkins University APL/Carnegie Institution of Washington Image 5770878 (https://tinyurl.com/yy96rrkr); 
Moon: NASA Image AS17-M-0831, hosted on the LPI website (https://tinyurl.com/y6mhpf7p); 
Mars: NASA/JPL/USGS Image PIA00425 (https://tinyurl.com/yyxnplfw).}
\label{Fig__Crusts}
\end{figure}

Figure \ref{Fig__Crusts}
condenses the present 
understanding on the 
composition, origin, and
associated volume of the
primary and secondary
crusts of Mercury, 
the Moon, and Mars.
\add{For Mercury, the secondary crustal
volume is based on the central value
of \citet{Padovan2015}
for the thickness of the crust,
$35\pm18$ km.
For the thickness of a potential
primary crust we use the value
of 10 m, based on the possible
range between 1 cm 
and 10 km obtained by 
\citet{VanderKaaden2015}.
The lower end member corresponds
to the lower measured lunar 
carbon content, while the
upper end member
corresponds to the
carbon content of the
carbon-rich CI meteorites
(Section \ref{Sec__PrimaryCrust}).
In calculating the volumes, a total 
silicate thickness of 419 km is assumed,
based on the 
results of \citet{Hauck2013} for the 
radius of the core and of \citet{Perry2015} 
for the planetary radius.
The thickness of the lunar crust comes
from \citet{Wieczorek2013}, while the
volume of secondary crustal material assumes
a total magmatic reservoir that 
is ten times the volume of the
extruded volcanic material 
as estimated in \citet{Head1992}.
Values for the volumes are based on 
a lunar radius of 1737 km \citep{Wieczorek06}
and a core radius of 330 km \citep{Weber11}.
The Martian crustal thickness range comes from
\citet{Plesa2018}, and volumes are based on a
core radius of 1850 km \citep{Plesa2018} and
a planetary radius of 3390 km 
\citep{Seidelmann2007}.}
Interestingly, \add{independent of the 
particular sources used in creating
Figure \ref{Fig__Crusts}, Mercury, the
Moon, and Mars,} despite 
their different sizes, 
thickness of the mantles,
and dominant crustal type,
have roughly 7 to 
10\% of their total silicates
contained in the crust.


\section{Impacts}\label{Sec__Impacts}

Craters are created from an impactor population,
which has a power-law size distribution
comprising relatively few very large objects
\citeg{Strom2015}.
These large members  are responsible 
for creating some of the largest impact structures 
observed on the 
surface of the terrestrial planets, such 
as the Caloris basin on Mercury
\add{\citep[basin diameter $D_{\rm b}\sim1550$ km,][]{Fassett12}}, 
the South-Pole Aitken basin on the Moon
\add{\citep[$D_{\rm b}\sim2200$ km,][]{Garrick-Bethell2009}},
and the Hellas basin on Mars 
\add{\citep[$D_{\rm b}\sim2070$ km,][]{Frey2008}}.
\add{Most impacts occur early on, and}
the \add{cumulative} number of impacts occurring
on a given surface unit will grow in time,
thus potentially providing a way to date
the unit based on the statistics of the 
crater population.
Indeed, crater statistics is the most powerful tool
to date the surfaces of solid solar system 
objects \citep[see][for a recent review]{Fassett2016}.
Comparison of crater statistics for different
units of the same object provides relative ages.
In the case of the Moon, absolute ages
can be obtained through calibration with
Apollo samples \citeg{Stoeffler2001}.
Then, the scaling of the impactor population to 
different parts of the solar system
provides an indirect way to assess the
absolute age of surficial units for planets
where only crater statistics is available
\citeg{Marchi2009}.




Outside of the limited range of impact
parameters available in a laboratory setting,
the details of impact processes can 
be simulated using
smoothed particle hydrodynamic codes
\citeg{Monaghan1992},
shock-physics codes like iSALE\footnote{\add{Access to the code can be obtained from https://isale-code.github.io/access.html}} 
\citep[e.g.,][]{Wuennemann2006}, 
or simple scaling laws \citeg{Pierazzo1997}.
Here, we focus on basin-forming impacts,
energetic events that have
the potential of interacting
with the temperature field of 
the mantle modifying locally 
its convective vigor 
\citep{Elkins-Tanton2004,Ghods2007,Padovan2017}.
These are smaller than planetary-scale 
events like the Moon-forming impact, 
but larger than the events that create
the smaller class\add{es} of impact craters
observed on the surfaces of airless 
rocky bodies (e.g, Moon, Mercury, large 
asteroids).
As a rule of thumb, impacts where
the size of the impactor is small 
with respect to the size of the
target (i.e., the planet) 
and that result in the formation of 
basins (large, multi-ring craters)
have the potential of interacting
with mantle convection.

The potential effects of impacts 
on surface magnetization and 
dynamo action in the
core are described in 
Section \ref{Sec__MagField}. 
Here we discuss the connection
between impacts and the thermal state
of the mantle,
with a focus on the potential surface
signatures of this interaction.
In so doing, we are explicitly making
a qualitative distinction between
impacts as a dating tool 
\citep{Fassett2016}, which is key in
deciphering the history of crustal
building (Section \ref{Sec__SecondaryCrust}),
and (large) impacts as a potential  
energy source of mantle dynamics.


\subsection{Basin formation}
The formation of an impact basin is a relatively 
fast process taking place on timescales of few
hours, even for the largest 
basins \citep[e.g.,][]{Potter2012}.
Schematically, an object hits the surface
of a planet at hyper-velocity, creating a
so-called transient cavity, which then collapses to 
form a shallower 
basin \citep{Melosh1989,Melosh2011}.
In this process, shock waves are produced
that travel in the interior, whose
energy is released as thermal energy
\add{\citep{Bjorkman1987}}
that increases the temperature
in a volume roughly centered 
along the vertical axis of
the contact point, possibly
causing large scale melting events.
The most widely-used code for
the accurate simulation of large impacts, 
events that cannot be reproduced
in a laboratory setting, is the
iSALE hydrocode \citeg{Wuennemann2006}.
Scaling laws aim at condensing in  
analytical expressions 
the main effects of impacts 
as observed in laboratory and 
numerical experiments.
They represent a relatively  
simple way to describe
how the properties of the impactor
(size, speed, composition) and of 
the target (composition, gravity)
control a number of physical quantities
that are key in the description
of the impact process.
In this section the goal is to 
elucidate the connections 
of large impacts with
the interior dynamics,
which occurs through the deposition
of the impact shock energy in the
subsurface.
Accordingly, we focus on the 
properties of the 
thermal anomaly that forms 
in the subsurface as a 
result of the release of the 
energy of the shock wave.

The computation of the 
impact-induced thermal
anomaly occurs through the 
following steps.
First, a connection is established
between the size of the observed
basin and the diameter and 
velocity of the impactor,
according to the following
scheme, where 
each arrow indicates a connection
through a scaling law: basin 
diameter $\rightarrow$ transient 
cavity diameter $\rightarrow$ diameter
and velocity of the impactor.
Estimates for the velocity of the 
impactor are based on statistical models
of the impactors population for the 
body of interest 
\citep{LeFeuvre2008,Marchi2009}.
The combination of these estimates 
with the size of the observed basin
provides the diameter of the impactor.
From the velocity and diameter of
the impactor,
additional scaling laws provide the
distribution of the shock pressure 
in the subsurface, whose release
is responsible for the impact heating,
i.e., for the release of energy
to the target \citep[e.g.,][]{Watters2009}.
The following equation highlights 
the key dependencies in the expression
of $H_{\rm imp}$, the energy 
release associated to a given impact:
\begin{eqnarray}
H_{\rm imp}=F\left(\rho_{\rm P},\rho_{\rm T},g, D_{\rm P}, v_{\rm i}\right),
\end{eqnarray}
where the gravity $g$ and density $\rho_{\rm T}$
of the target are known.
The projectile (i.e., impactor) 
density $\rho_{\rm P}$ is unknown, while 
its diameter $D_{\rm P}$
and impact velocity $v_{\rm i}$ are
related to the basin diameter 
as described above.
To compute $H_{\rm imp}$, a number 
of assumptions are usually made.
First, despite impacts
occurs preferentially at $45^{\circ}$ 
from the vertical \citep{LeFeuvre2008}, 
only the vertical 
component of the velocity is considered, 
since it accounts for the majority
of the impact-related
effects \add{\citep{Shoemaker1962,Shoemaker1983}}.
Secondly, it is often assumed
that the impactor has the same
composition of the target (i.e., rocky),
which can be justified on the
basis that objects from 
the asteroid belt
are the main current contributors
to the Earth impactors population
\citep[e.g.,][]{Morbidelli2002}.
Both assumptions can be relaxed,
and studies exist on the 
effects of composition and
obliquity on the impact
processes \citeg{Elbeshausen2011,Ruedas2018}.
A recent and clear description of the computation
of the energy release for a given impact
can be found in, e.g., \citet{Roberts2017},
which accounts for the crater scaling laws
\citep{Schmidt1987,Melosh1989,Holsapple1993},
the shock-pressure scaling of 
\citet{Pierazzo1997}, and the 
foundering shock-heating method
of \citet{Watters2009}. 
Care should be taken in using scaling laws
appropriate for the problem at hand. 
For example, the pressure scaling
of \citet{Pierazzo1997} should only
be used for impactor velocities
$\gtrsim$ 10 km/s.
The scalings of \citet{Monteux2015},
based on iSALE-2D simulations, cover
the slow impactor range ($v_{\rm i}\gtrsim4$
km/s).

Given the very different timescales of 
basin formation processes (few hours) and
mantle convection (millions of years),
the use of scaling laws represented,
until recently, the only viable approach
to investigate the interplay of these
two processes \citeg{Roberts2012,Roberts2017}.
However, the increase in the computational
power available allows one to use outputs of
the short timescales simulations (such as 
those of iSALE) within the long timescales
of convection simulations 
\add{\citep{Rolf2016,Golabek2018}}.

\subsection{Impact-induced melting}\label{Sec__Impact-InducedMelting}

There are three types of melting
processes that can be induced by a large impact.
First, the release of the impact shock 
energy increases the 
temperature of a volume of the target 
around the impact location, possibly 
melting it completely \citeg{Roberts2012}. 
This process results in what is commonly 
referred to in the literature as 
a melt pond or melt sheet
\citeg{Cassanelli2016}.
Second, during the formation of a basin 
a transient crater is excavated,
whose maximum depth is larger than
the depth of the final 
basin \citeg{Melosh1989}.
For very large events this transient 
crater can reach depths of 
tens of kilometers, thus locally
depressing the solidus and potentially
inducing massive in-situ 
decompression melting.
This possibility was initially 
suggested by \citet{Ronca1966}---the
seminal work on the connection of
impacts and volcanism---and was
subsequently
discussed several times in relation
to the potential connection of 
impact events with the generation
of large igneous provinces
\citeg{Rogers1982,Rampino1987,
Elkins-Tanton2005}.
Third, the positive thermal anomaly
resulting from a large impact event
locally warps the isotherms producing
dynamical currents in the mantle
under the newly formed basin,
which could lead to additional 
post-impact convective 
decompression melting
\citeg{Grieve1980}.

The solidification of the melt sheet
could occur through equilibrium
crystallization, which preserves the bulk
composition of the melt, or through
igneous differentiation, which would
induce a compositional gradient in the
solidified melt sheet
\citep{Cassanelli2016} (see Section \ref{Sec__MagmaOceans}).
According to the model of 
\citet{Cassanelli2016},
the discrimination between the
two possibilities is controlled by
the size of the crystals that form
in the cooling pond, while the bulk
size of the melt sheet plays a 
minor influence. 
However, the crystal size is difficult
to determine a priori.
On the Moon,
there are indications for 
igneous differentiation
in the Orientale basin 
\citep{Spudis2014} and 
in the South-Pole Aitken 
basin \citep{Vaughan2014}.

The process of in-situ
decompression melting is predicated 
on the assumption that in the short
timescales of transient crater 
formation rocks at depth undergo melting
due to the rapid removal of
lithostatic pressure.
However, \citet{Ivanov2003}
argue that a more appropriate estimate is 
based on the profile of the final crater,
which, being shallower than the
transient crater, would 
induce significantly less melting.
Independent of the feasibility of the
in-situ decompression melting process,
it is important to note that 
both the solidification of the melt sheet
and the eruption of the 
in-situ decompressed melt
would occur on geologically short
timescales \citep{Reese2006,
Elkins-Tanton2004,Elkins-Tanton2005}.
Thus, dating of units related to
the first two processes would 
coincide with the age of the
basin itself.

\subsection{Impact-related effects on global mantle dynamics}

The observations that most basaltic
units on the near side of the Moon
are located within large impact basins
and have been emplaced from several tens
to up to hundreds of millions of years
after the basins formation,
hinted at a possible
causal connection between impacts
and subsequent volcanism.
This process would result from 
the interaction of the impact-induced 
thermal anomaly with the 
mantle temperature field, the process
referred to above as 
post-impact convective 
decompression melting.
While a qualitative 
assessment of post-impact
convective decompression melting
has been published early 
on \citep{Grieve1980},
more quantitative work has appeared
only in the last 20 years.

\citet{Reese2002,Reese2004}
showed that a thermal anomaly 
induced by a very ancient impact event 
can explain the long-term volcanism 
observed in the Tharsis province on Mars
\citep{Hartmann1999,Werner2009}.
However, conventional thermal 
evolution models do not require a large 
impact to justify the long-term 
volcanism on Tharsis, 
and are also compatible with 
several additional constraints
\citeg{Plesa2018}.
More recently, 
\citet{Roberts2012b,Roberts2017}
investigated the role of large impacts 
on the dynamics of the mantle and
energetics of the core of
Mars. 
While large impacts can focus
upwellings under the impact locations
\citep{Roberts2012b},
a confirmation of the results of
\citet{Reese2004},
and can temporarily frustrate dynamo
action in the core, the duration
of a Martian dynamo is 
controlled by the rheological
properties of the lower 
mantle \citep{Roberts2017}.
Similar models for Mercury, where
the largest confirmed
impact basin is the relatively 
small Caloris basin, with 
a diameter of about 1500 km
\citep{Fassett2009},
show that the effects on the 
energetics of the core are minor,
while inducing a modification of
the convection planform under 
the basin location \citep{Roberts2012}.
\citet{Rolf2016} investigated the
effects of the lunar impact history
on the long-term evolution of the 
body, concluding that the full
sequence of large impacts
may influence the 
present-day heat flux
and the initial expansion
of the Moon, two parameters
that could be measured and thus,
provide additional 
constraint for thermal evolution
models 
(Section \ref{Sec__AdditionalConstraints}).
However, 
\citet{Miljkovic13} proved that
the different morphologies of the
lunar impact basins bewteen the near-side
and the far-side can
be explained by the asymmetric
distribution of heat producing elements 
in the interior 
before the basins formed 
\citep{Laneuville2013,Laneuville2018}.
Accordingly, it seems plausible,
at least for the Moon, that 
\add{while large impacts may
induce effects that are still measurable 
today \citep{Rolf2016,Ruedas2019}, it is the
initial distribution of heat-producing
elements that largely controls the overall
evolution.}

\subsection{Local signatures of impact-related effects}
The studies described in the
previous section all dealt with
the potential global 
consequences of one or more 
large impacts.
However, the creation of an impact basin
is by definition a local process,
which ejects or melts pre-existing crust
\citeg{Potter2012}, thus
modifying the local surficial composition
and topography.
Often, large basins on the Moon and
Mercury are associated with
the presence of positive gravity anomalies 
(or mascons, for
mass concentrations) and
volcanic infillings, which on the 
Moon are called maria \citep{Melosh2013,James2015}. 

\citet{Elkins-Tanton2004} investigated the 
potential causal connection of 
impact processes with subsequent
localized enhanced volcanism.
They developed a model for the
magmatic effects of large impacts, which 
takes into account the 
in-situ decompression melting and
the post-impact convective decompression
melting (see Section \ref{Sec__Impact-InducedMelting}
for definitions).
By adjusting the pre-impact lunar mantle 
temperature profile,
\citet{Elkins-Tanton2004} were able to reproduce 
the source depth and volume of the maria.
However, it was difficult to match the interval
observed between the formation of the basin
and the later emplacement of the volcanic material.
The authors recognize that 
a potential shortcoming of the model
lies in their assumption that 
the mantle, before an impact,
is not convecting. 
Accordingly, any post-impact convective
decompression melting is only due to 
the limited convection activity induced 
by the warping of the isotherms---and 
related horizontal temperature 
gradient---induced by the impact itself.


\add{Using thermal evolution
models that include two-phase flow,
\citet{Ghods2007} investigated convective
long-term magmatic activity
induced by lunar impact basins
using as background temperature
field a set of models evolved \addd{for 500 Myr},
when the largest impacts are 
assumed to occur.
This model is able to reproduce the timing
of volcanic activity within Imbrium-sized 
basins and, for certain combination of 
parameters, the observed volume of
volcanic material. 
As for the case of \citet{Elkins-Tanton2004},
the model predicts a large amount of 
volcanic activity within the large SPA basin,
which is not observed.}

\begin{figure}[t!]
\centering
\includegraphics[width=0.8\textwidth]{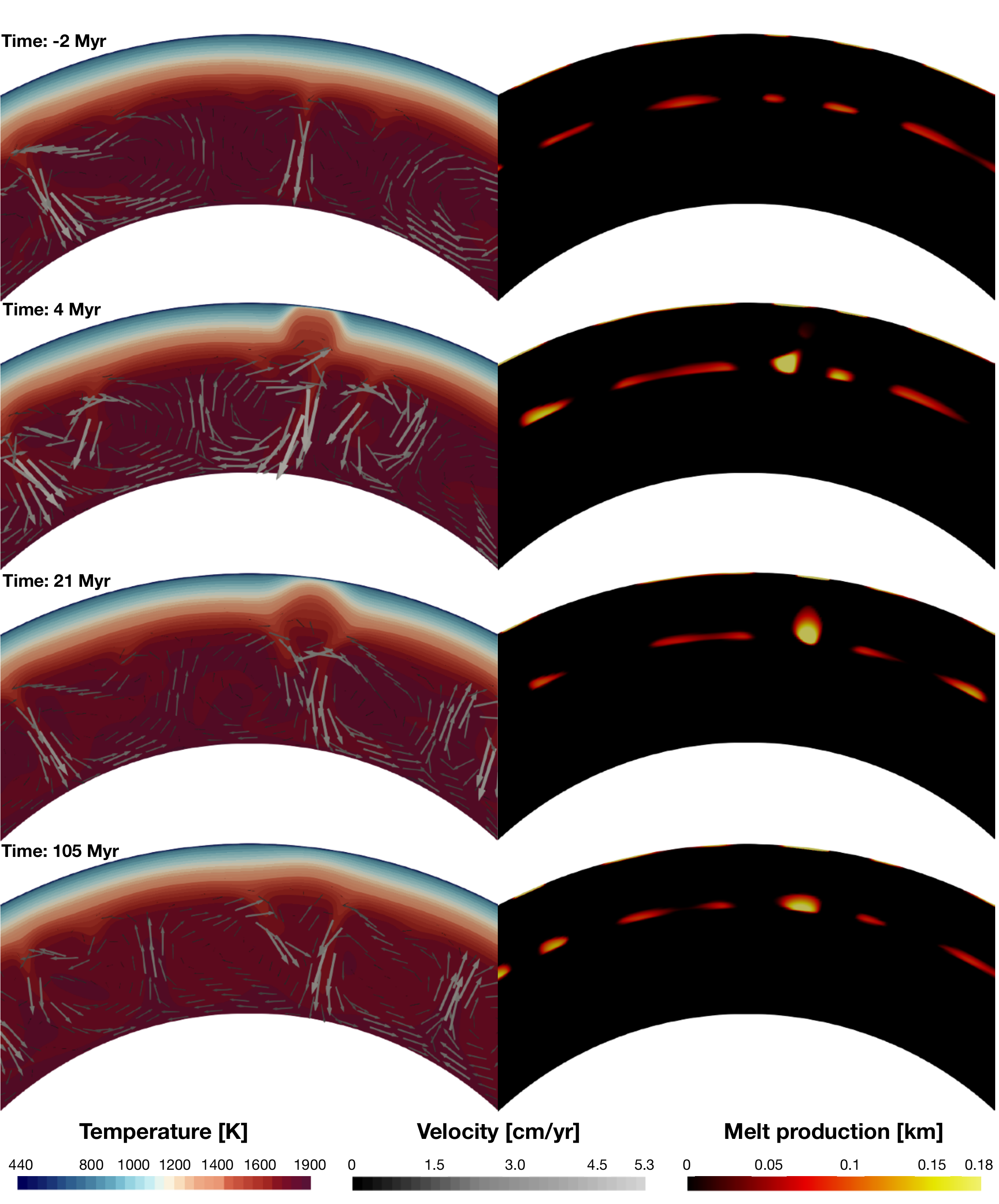}
\caption{Formation of a Caloris-sized basin on Mercury. 
Each panel indicates the timing with respect to the basin-forming
impact. 
Left column: Temporal evolution of the temperature field (background) and of the magnitude of the velocity field
(arrows). Right column: temporal evolution of the melt production, which is expressed in equivalent crustal thickness. 
The thermal anomaly warps the isotherms and 
convective currents, which generate melt, 
develop in the region close to the impact,
but with a certain delay 
with respect to the impact event. \add{Redrawn after}  
\citet{Padovan2017}.}
\label{fig:impacts}
\end{figure}

\citet{Padovan2017} 
investigated the magmatic effects
of the formation of large impact basins
on Mercury
using as background temperature and velocity
field a reference model selected from
a large suite of 2D dynamical models.
The reference model was selected based
on its compatibility with the observed thickness
of the crust \citep[as inferred by 
geoid-to-topography ratios, ][]{Padovan2015},
the duration of volcanic activity 
\citep[as inferred by crater statistics,][]{Byrne2016}, and
the depth of the lithospheric thickness at two 
different epochs \citep[as inferred by petrologic
experiments,][]{Namur2016}.
Figure \ref{fig:impacts} illustrates how the 
formation of a Caloris-sized basin produces
a large thermal anomaly in the mantle, which
locally alters the magmatic activity for a 
geologically long timescale.
By tracking the amount of melt produced under
a basin, \citet{Padovan2017} were able to
reproduce the volume and time of emplacement
of the volcanic materials within the Caloris
and Rembrandt basins on Mercury.
This result critically depends on the 
thermal state of the mantle at the time
of the basin formation, thus indicating
how basin-specific datasets (namely, volume
and time of emplacement of volcanic material)
provide information on the global evolution
of the body.

\section{Mantle cooling and magnetic field generation}\label{Sec__MagField}

Besides the Earth, among the solid planets and satellites of the solar system only Mercury and Ganymede possess today a self-sustained magnetic field. 
In addition, the surfaces of Mercury, the Moon, and Mars exhibit a remnant magnetization due to the presence of magnetic minerals in their crust and lithosphere. On the one hand, an active magnetic field requires organized, large-scale motion of an electrically conducting fluid, generally thought to be the liquid core. On the other hand, a remnant magnetization, detected via remote sensing or measured by direct sample analysis, indicates that a dynamo was active in the past so that the accompanying magnetic field left its imprint in surface and shallow rocks when these formed. The possibility to generate a magnetic dynamo is tightly related to the way mantle convection operates by cooling the core \citep{Stevenson83}. Therefore, the existence of an internally-generated magnetic field, at present or in the past, can be used as an indicator of convection and cooling of the mantle and as a constraint for its thermal evolution.

Ferromagnetic minerals present in crustal rocks acquire a so-called thermoremanent magnetization (TRM) if cooled from above their Curie temperature ($T_C$) in the presence of an ambient magnetic field. 
Melts that erupt on a planetary surface or are intruded at depth typically have temperatures largely in excess of $T_C$ and, in the presence of a background magnetic field, acquire a strong TRM upon cooling below $T_C$. If the age of the magnetized rocks is also known---from radiometric techniques or from dating of surface units---it is possible to infer the lifetime of an ancient magnetic field and place constraints on the thermal history of the body.

The generation of a planetary-scale magnetic field is possible in the presence of a large volume of electrically conducting fluid whose motion is driven by an energy source---thermal, compositional, and/or mechanical---and structurally organized through rotation, which provides the positive feedback necessary for the field to be self-sustained \citeg{Olson2008}. 
Although promising advances in theory and experiments indicate that core motion and turbulence can be excited by periodic forcing induced by tides, precession, and libration \citep{LeBars2016}, buoyancy-driven flows of thermal and/or chemical origin provide today the standard model of core convection and dynamo action \citep[see, e.g. the review by][]{Breuer2010}. 

Thermal and compositional convection in a liquid core are driven by mantle cooling. From a global-scale perspective, thermal convection requires that the average heat flux at the base of the mantle ($F_\text{CMB}$) exceeds the heat flux conducted along the adiabat of the convective core ($F_\text{ad}$), i.e., 
\begin{equation}
    F_\text{CMB} = k_\text{m}\frac{dT_\text{CMB}}{dr} \geq F_\text{ad} = k_\text{c}\frac{\alpha_\text{c} g_\text{c} T_\text{CMB}}{c_\text{c}}, \label{eq:Fcmb}    
\end{equation}
where $r$ is the radial coordinate, $k_\text{m}$ and $k_\text{c}$ the thermal conductivities of the mantle and core, $\alpha_\text{c}$ the coefficient of thermal expansion of the liquid core, \add{$g_\text{c}$ the gravity acceleration at the CMB}, $T_\text{CMB}$ the temperature at the CMB, and $c_\text{c}$ the core specific heat capacity. A large temperature gradient at the base of the mantle helps satisfying this condition. On Earth, the subduction of tectonic plates favours cooling of the interior; it leads to a large temperature difference between the mantle and core and in turn to a high heat flux at the base of the mantle, which facilitates core convection and cooling over Earth's history \citep[e.g.,][]{Nakagawa2015}. In contrast, on stagnant lid bodies, the interior is cooled less efficiently; the mantle tends to remain hotter and the temperature gradient at the CMB smaller, with the consequence that driving thermal convection in the core is generally possible only during the early stages of evolution if the process of core formation left the core largely superheated with respect to the mantle \citep[e.g.,][]{Breuer2003,Williams2004} and/or upon the first mantle overturn corresponding to onset of thermal (and possibly compositional) mantle convection \citep[][]{Elkins-Tanton2005c,Plesa2014}. In addition, the presence of heat-producing elements such as K, as well as the release of latent heat upon core freezing (see below) can also contribute to drive thermal convection in the core.
    
For compositional convection to occur, the core must be 1) composed of a mixture of Fe and one or more light alloying elements---S being the best-studied candidate---and 2) cooled sufficiently to start local freezing. The presence of light elements reduces the density of the core alloy and its melting temperature with respect to those of a core with a pure Fe composition. 

Let us consider a standard model of freezing of a Fe-S core. If the pressure slope of the core melting temperature is larger than that of the core adiabat as in the case of the Earth, upon cooling of a fully liquid core, the temperature drops below the liquidus first at the center of the planet, causing the core to freeze from the bottom up in a similar way as a magma ocean would freeze (Figure \ref{fig:overturn}a). As cooling and crystallization continue, the liquid outer core becomes more and more depleted in Fe and enriched in S. Atop of the solidified core, a sulphur-rich layer forms. Being lighter than the overlying fluid, this layer buoyantly rises driving compositional convection that tends to homogenize the remaining fluid and can potentially power a dynamo.

In the framework of the Fe-S system, a different crystallization scenario is also possible. Melting experiments indicate that at pressures smaller than $\sim$14 GPa---relevant for the cores of Mercury, the Moon, and Mars---the pressure slope of the core melting temperature is negative \citep{Chen2008,Buono2011}. As a consequence, the adiabatic temperature of a cooling core would first  drop below the liquidus at the CMB rather than at the center. Fe crystals would then form at the CMB and precipitate, generating an ``iron-snow'' \citep[see e.g.][for a review]{Breuer2015b}. Since the melting temperature of an Fe-S liquid is lower than that of pure Fe, newly formed Fe crystals would rapidly remelt, enriching the remaining fluid in Fe. Dense, Fe-rich fluid atop lighter Fe-FeS fluid creates a gravitationally unstable configuration that triggers compositional convection and possibly dynamo action \citep[e.g.,][]{Rueckriemen2018}. 

\subsection{Mercury's present-day and early magnetic field}


The current knowledge of Mercury's magnetic field is largely the result of more than four years spent by the MESSENGER spacecraft orbiting the planet \citep[see][for a recent review]{Johnson2018}. The field is weak, axisymmetric, dipolar, and characterized by a large quadrupolar component corresponding to a northward offset of the magnetic equator with respect to the geographic equator of about 20\% of the planetary radius \citep{Anderson2012}. Taken together, these features pose a significant challenge for numerical dynamo models \citep{Wicht2014}. On the one hand, standard models of core flow driven by compositional convection associated with an Earth-like growth of the inner core can produce under certain conditions a weak magnetic field, but not other key characteristics such as the small amplitude of harmonic components greater than 2 and the large northward offset of the dipole \citep{Heimpel2005,Stanley2005,Takahashi2006}. On the other hand, models involving a stably stratified (i.e. non-convecting) layer at the top of the liquid core, with convection and dynamo action taking place beneath it, successfully predict a low-amplitude field with large dipolar and quadrupolar components \citep{Christensen2006,Christensen2008,Takahashi2019}. 

Core flow driven by a laterally-variable heat flux extracted at the base of the mantle has been identified as a potentially important ingredient to induce a stable dipolar field with a large quadrupolar component \citep{Cao2014,Tian2015}. Because of its 3:2 spin-orbit resonance, Mercury experiences an uneven insolation that results in large-scale differences of its surface temperature with cold poles at $0^\circ$ and $180^\circ$ latitude, and equatorial hot and warm poles at $0^\circ$ and $180^\circ$ and $\pm90^\circ$ longitude, respectively. Such pattern diffuses through the mantle leading to a high CMB heat flux at the poles (where the mantle is cooler) and a low heat flux in the equatorial regions (where the mantle is hotter) \citep{Tosi2015}. \citet{Cao2014}, however, found that the characteristics of Mercury's magnetic field are best reproduced when assuming the highest heat flux to be at the equator. Although such distribution could be important to constrain the planform of mantle convection, no obvious mechanism seems to be able to generate it and it is at odds with the heat flow pattern expected on the base of the surface temperature. The same holds for the degree-2 pattern postulated by \citet{Tian2015}. The authors obtained a Mercury-like magnetic field assuming a high heat flux in the northern hemisphere, which they posited to be a remnant of the mantle activity that produced the northern volcanic plains 3.7--3.8 Ga (see Section \ref{Sec__Crust}). However, no mechanism has so far been proposed to generate hemispherical volcanism on Mercury and, at any rate, maintaining the required heat flux throughout Mercury's evolution seems unlikely. Overall, a conclusive dynamo theory for Mercury and its link to the cooling mantle are yet to be established.

\begin{figure}[t!]
\centering
\includegraphics[width=0.55\textwidth]{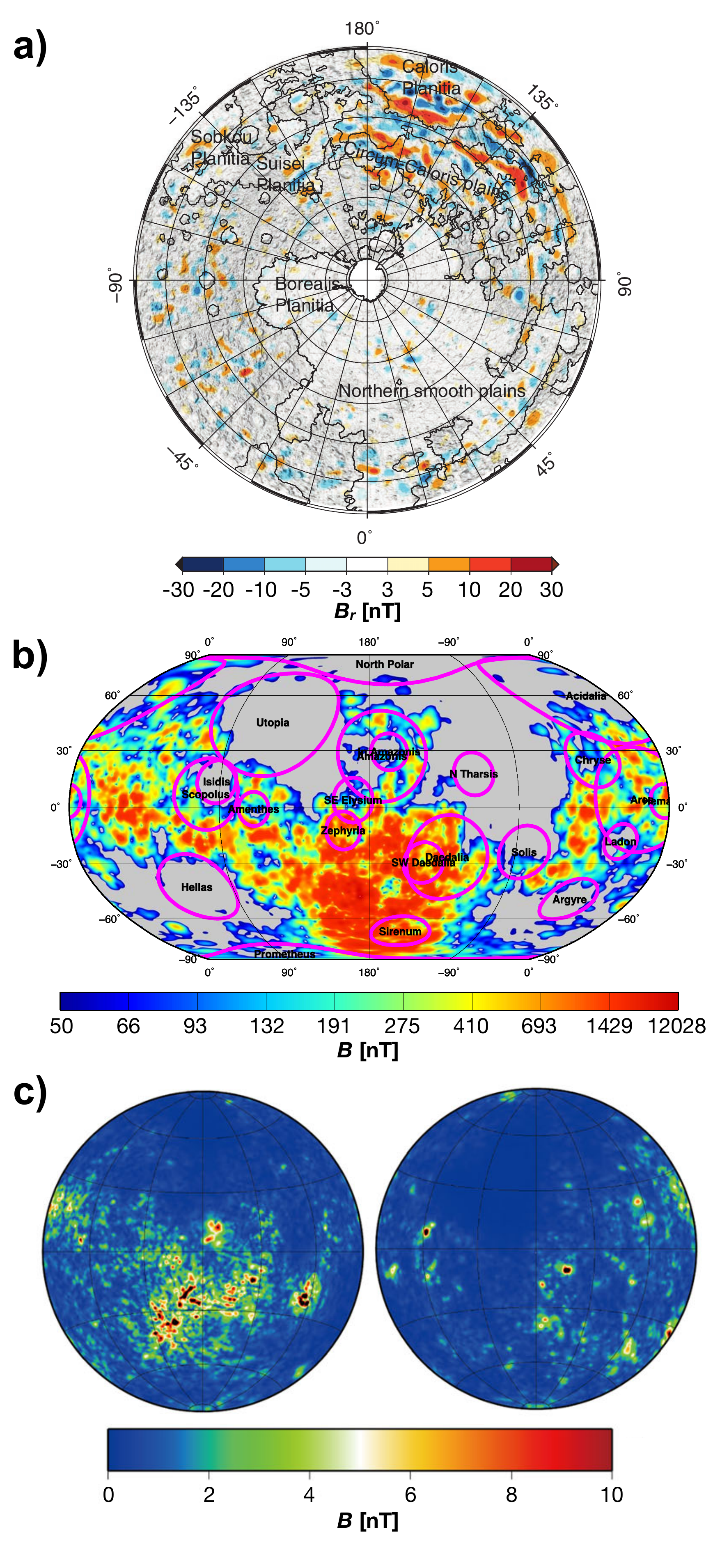}
\caption{a) Map of the radial component of Mercury's magnetic field at 30 km altitude based on a model of lithospheric magnetization (from \citet{Johnson2018}, Figure 5.19). b) Total intensity of the lithospheric magnetic field of Mars evaluated at the mean planetary radius of 3393.5 km.  Thick purple circles indicate major impact basins (from \citet{Morschhauser2016}, Figure 10.8). c) Total intensity of the lithospheric magnetic field of the lunar nearside (left) and farside (right) at 30 km altitude (from \citet{Tsunakawa2015}, Figure 2).}
\label{fig:magfield}
\end{figure}
	
A low-altitude observation campaign carried out between 2014 and 2015 during the final phase of the mission, allowed MESSENGER to probe small-scale, low-amplitude magnetic fields of crustal origin. \citet{Johnson2015} identified a field with a peak intensity of about 20 nT 
in the Suisei Planitia region, consistent with the TRM of the underlying crust (Figure \ref{fig:magfield}a). The analyzed region encompasses different geological units and, in particular, smooth volcanic plains emplaced \addd{between} 3.9 and 3.7 \addd{Ga}. Subsequent analyses revealed crustal magnetization associated with the Caloris and circum-Caloris region, which is of similar age \citep{Hood2016}. Overall, these observations indicate that a dynamo-generated field, most likely with a strength between that of the present-day field and up to a factor hundred higher (i.e. similar to the Earth’s) was active on Mercury at least until 3.7 Ga \citep{Johnson2018}. Models of the interior evolution of Mercury need thus to be compatible with an ancient dynamo active around 3.7 Ga as well as with the present-day one.

Whether or not a dynamo was active continuously throughout the planet’s history is unknown and represents a major source of uncertainty. Thermal evolution models of the interior of Mercury 
predict an initial phase of efficient crust production where the mantle heats up because of radiogenic heating \citep[e.g.,][]{Tosi2013,Padovan2017, Hauck2018}. During this phase, the temperature difference between the mantle and core decreases and so does the heat flux at the CMB, which rapidly drops below the critical adiabatic value. A sub-adiabatic heat flux at the CMB---a common feature throughout most of the evolution of one-plate bodies---causes the top of the liquid core to become thermally stratified and suggests that thermal convection driven by mantle cooling alone can not be responsible for maintaining a dynamo over long time spans. 

As for the cores of the other terrestrial planets, the composition of Mercury's core is not well known. \citet{Dumberry2015} showed that under the assumption of a Fe-S composition, a present-day magnetic field together with geodetic constraints on Mercury's moment of inertia and rotation state can be best accounted for if the inner core is relatively small ($\leq 650$ km) and operates in the Fe-snow crystallization regime.
However, a Fe-Si composition is thought to be more relevant for Mercury than a Fe-S one \citep{Knibbe2018}. The low FeO content of the surface \citep{Nittler2011} hints at reducing conditions for the mantle \citep{Zolotov13} which, during planet formation and core segregation, may have favoured the fractionation of Si rather than S in the core  \citep[e.g.,][]{Malavergne2010}. In a crystallizing Fe-Si core, the amount of Si fractionating between solid and liquid is small \citep{Kuwayama2004}. Upon freezing, negligible compositional buoyancy is generated, which thus can not drive core convection and dynamo action. In contrast to previous efforts largely based on the assumption of a Fe-S composition \citep{Grott2011,Tosi2013,Dumberry2015}, \citet{Knibbe2018} developed a model of Mercury's interior evolution including the crystallization of a Fe-Si core and, importantly, the self-consistent formation of a thermally-stratified layer at the top of the liquid core caused by the expected drop of the mantle heat flux at the CMB in the early evolution. The authors showed that under these circumstances, 
core convection and an accompanying magnetic field can be powered by the release of latent heat (higher for a Fe-Si core than for a Fe-S one) within a thin liquid shell located between a liquid, but thermally-stratified, upper core, and a growing solid inner core. 
In agreement with the evidence of remnant crustal magnetization \citep{Johnson2015}, the model of \citet{Knibbe2018} also predicts an early onset of magnetic field generation, which is expected to continue until present due to the slow growth of the inner core.

    
\subsection{Mars' early magnetic field}

At present, Mars does not possess a global-scale magnetic field of internal origin, but exhibits a strong, small-scale crustal magnetization largely associated with the old, heavily cratered southern highlands \citeg{Acuna1999} (Figure \ref{fig:magfield}b). Therefore, a dynamo-generated field must have existed in the past. The lack of significant magnetization within large impact basins can be used to infer the time at which the Martian dynamo stopped operating. Impacts can reset the surface age as well as surface and crust magnetization over an area roughly corresponding to the size of the resulting crater \citep[e.g.,][]{Mohit2004}. The age of weakly- or non-magnetized basins provides thus an upper bound on the age of the dynamo and in turn a constraint for the thermal evolution of the interior. Analyses of the distribution of magnetization (or the lack thereof) of large basins 
together with the corresponding ages 
indicate that a dynamo stopped to be active on Mars before $4.0-4.1$ Ga  \citep[e.g.,][]{Acuna1999,Lillis2008,Lillis2013}. 

Most models of the thermal evolution of Mars' interior, which can satisfy the above requirement of an early dynamo, consider this to be of thermal origin, i.e., powered by thermal convection in the core driven by a super-adiabatic CMB heat flux \citep[e.g.,][]{Schubert1990,Hauck2002,Breuer2003,Williams2004,Morschhauser2011,Samuel2019}, although more speculative hypotheses such as that of a tidally-driven dynamo also exist \citep{Arkani-Hamed2009}. The required heat flux in excess of the core adiabatic gradient (eq. \eqref{eq:Fcmb}) is easily obtained during the early evolution---for Mars as well as for other bodies---by simply considering the core to be initially super-heated with respect to the mantle, an expected outcome of core segregation \citep{Solomon1979}.
This simple scenario is particularly plausible for Mars, which accreted and differentiated within the first few million years of the solar system \citep[e.g.,][]{Nimmo2007,Dauphas2011}, a compressed timescale that likely prevented the interior from achieving thermal equilibration and favoured instead a large temperature difference between core and mantle.

Building on an early hypothesis of \citet{Sleep1994}, \citet{Nimmo2000} argued that a strong heat flux at the CMB may have also been caused by cooling of the core due to plate tectonics operating during the first $\sim$500 Myr of Mars' history and followed by a stagnant-lid regime that persisted until present. The reported discovery from Mars Global Surveyor data of alternating magnetic lineations in the southern hemisphere resembling those that characterize spreading oceanic ridges on Earth \citep{Connerney2005} provided some observational ground to the hypothesis of early plate tectonics. However, the efficient mantle cooling initially caused by sinking plates tends to delay the subsequent production of crust once the stagnant lid regime is established because the sub-lithospheric mantle needs a relatively long time (1--2 Gyr) to heat up and produce partial melt \citep{Breuer2003}. A delayed crust production is at odds with the evidence that volcanism and crust production on Mars were concentrated in the Noachian period \addd{(about 4.5 to 3.7 Ga)} and declined since its end  (see Section \ref{Sec__Crust}), thus making the plate tectonics hypothesis somewhat problematic.

While the generation of a thermal dynamo necessarily requires a sufficiently high CMB heat flux, its cessation may also be the result of large impacts \citep{Roberts2009,Roberts2012b,Roberts2017,Monteux2015} 
The release of shock-pressure associated with impacts locally raises the mantle temperature and promotes the formation of upwellings beneath the impact region (see Section \ref{Sec__Impacts}). As a consequence, the average heat flux at the base of the mantle tends to decrease. Several giant impacts likely occurred on Mars within a narrow time window ($\sim$100 Myr), around the time at which Mars' global magnetic field disappeared \citep{Frey2008,Frey2013}. Simulations of mantle convection including the (cumulative) thermal effects of impacts have shown that for events generating basins larger than 2500 km, 
the accompanying decrease of the CMB heat flux can be sufficient to stop the dynamo \citep{Roberts2012b}. 
However, whether the dynamo is halted only temporarily or permanently, ultimately depends on the thermal state of the mantle preceding the impact as well as on its viscosity and thermal conductivity \citep{Roberts2012b,Roberts2017}, thus making it difficult to robustly evaluate this hypothesis.


\subsection{The long-lived lunar magnetic field}

Like Mars, the Moon lacks a present-day magnetic field, but it has been known since the Apollo era that lunar rocks and crust are magnetized \citep[see, e.g.][]{Fuller1987}. Laboratory analyses of numerous Apollo samples show evidence of a paleomagnetic field from 4.25 Ga until possibly as recently as 200 Ma \citep[see][for a review]{Weiss2014}. However, the inferred field intensity varies widely among samples of different ages. Until 3.56 Ga, samples show a mean paleointensity of several tens of $\mu$T \citep[e.g.,][]{Garrick2009,Shea2012,Suavet2013}.
The intensity declines abruptly by about one order of magnitude around $3.2-3.3$ Ga and remains at the level of few $\mu$T for younger samples. It has been recognized that due to the limited accuracy associated with the retrieval of the TRM of weakly magnetized samples, the low intensities inferred for samples younger than 3.3 Ga are actually compatible with a vanishingly small field \citep{Tikoo2014,Buz2015}. However, the recent measurements 
by \citet{Tikoo2017} appear sufficiently robust to indicate that a weak field of 5 $\mu$T was indeed present at least 2.5 Ga, more likely around 1 Ga, firmly extending the lifetime of the lunar dynamo by about 1 to 2.5 billion years (based on the previous idea that 3.56 Ga was the youngest robust age of the dynamo). The actual time at which the dynamo ceased to operate remains unclear.

Accounting for a long-lived magnetic field of widely variable intensity poses a severe challenge for dynamo models. No single mechanism is able to account for an early, intense, and relatively short-lived dynamo between 4.25 and 3.56 Ga and a  weak, long-lived one operating until (at least) 2.5 Ga and possibly longer. Several authors studied the conditions required to generate a magnetic dynamo on the Moon in the framework of coupled models of core-mantle evolution \citep[e.g.,][]{Konrad1997,Stegman2003,Evans2014,Laneuville2014,Scheinberg2015,Evans2018}. A simple thermal dynamo induced by the cooling of a superheated core is generally possible only during the first few hundred million years of evolution for a chemically homogeneous mantle \citep[e.g.,][]{Konrad1997,Laneuville2014}. A heat flux at the base of the mantle in excess of the core adiabatic heat flux can be maintained for a much longer time, even \addd{up to} 2.5 Ga, in the presence of a compositionally-stratified mantle following the overturn of a crystallized magma ocean, particularly for a hydrous lower mantle where, upon overturn, water, behaving as incompatible, is sequestered within the deepest cumulates \citep{Evans2014} (see Section \ref{Sec__MagmaOceans}). Considering, in addition to thermal buoyancy, the compositional buoyancy associated with the release of S upon core freezing (S is so far the only light alloying element considered in detail), easily extends the dynamo lifetime to billions of years \citep{Laneuville2014,Scheinberg2015}. A dynamo powered by core crystallization could even persist until present day unless a change in the regime of core crystallization from bottom-up to top-down takes place due to the progressive increase of S in the outer core \citep{Laneuville2014,Rueckriemen2018}. 

Overall, these models show that the generation of a long-lived lunar dynamo can be explained in terms of the well-known mechanisms of thermal and compositional buoyancy in the core induced by convective mantle cooling.
However, accounting for the inferred intensity of the resulting magnetic field\add{---both for the early and the late dynamo---}is much more difficult. This is usually computed on the base of scaling laws derived from numerical simulations that express the intensity of a dynamo-generated magnetic field at a planetary surface in terms of, among other parameters, the heat flux at the CMB \citep{Christensen2010}. 
As discussed by \citet{Evans2018}, \add{in principle, thermo-compositional convection in the core could power until a recent past a weak field of $\leq 1.9$ $\mu$T, not exactly as high as the 5 $\mu$T inferred by \citet{Tikoo2017}.
However, it cannot provide enough energy to generate a magnetic field of tens of $\mu$T until 3.56 Ga, even under the most optimistic assumptions on poorly known parameters.} 
Non-convective dynamos generated via mechanical stirring of the liquid core induced by precession of the Moon's spin axis \citep{Dwyer2011} or other instabilities associated with precession, librations, or tides that could be triggered by large impacts \citep{LeBars2011} also fall short by about one order of magnitude in predicting the surface intensity of the lunar magnetic field. A higher intensity of up to few tens of $\mu$T and hence close to the observed values, as well as a long-lived field could be accounted for by a ``silicate dynamo'' \citep{Ziegler2013} generated in a basal magma ocean atop the lunar core \citep{Scheinberg2018}. As discussed in Section \ref{Sec__MagmaOceans}, upon overturn of a crystallized magma ocean, dense cumulates enriched in incompatible HPEs may sink at the base of the mantle \citep[e.g.,][]{Yu2019} and heat up \citep{Stegman2003,Zhang2013}. If their density and heat sources content are sufficiently large, a basal magma ocean can form \citep[see e.g.][for a description of this process in the context of early Mars]{Plesa2014}. Provided that the poorly known, high-temperature electrical conductivity of the overturned silicates is sufficiently large, \citet{Scheinberg2018} showed that the surface intensity of a magnetic field generated in a thick basal magma ocean 
could be marginally compatible with the observations, largely because the surface strength of the magnetic field is positively affected by the vicinity and large size of the convective region. 

In synthesis, observational constraints provide a relatively clear picture of the evolution of the lunar magnetic field. Its interpretation, however, is non-trivial. No single mechanism of magnetic field generation is able to account for both the amplitude and the timing of the dynamo. A combination of different mechanisms such as thermal convection (in the core or in a magma ocean), compositional convection, or mechanically-driven instabilities acting at different times and possibly in concert could be key to explain the history of the lunar magnetic field in its entirety.

\section{Additional surficial manifestation}
\label{Sec__AdditionalConstraints}

We devote this section to a brief discussion of 
 additional constraints 
that can be placed on the interior 
evolution of stagnant-lid bodies.
\add{These constraints are  
both heavily model-dependent and their bearing 
on the interior evolution is somewhat indirect.
They include }
planetary contraction and expansion, surface heat flux, and thickness of the elastic lithosphere.

\subsection{Global contraction and expansion}

The thermal evolution of terrestrial bodies can be manifest in the geological record through extensional and contractional features such as grabens and thrust faults, which if interpreted as a global response to the changes in the overall thermal state of the body, testify to periods of warming and cooling, respectively. In fact, the  radius change $\Delta R$ of a planet due to thermal expansion and contraction can be expressed as \citeg{Solomon1976}:
\begin{equation}
    \Delta R = \frac{1}{R_\text{p}^2} \int_{0}^{R_\text{p}} \alpha \Delta T(r) \, r^2\, dr,
\end{equation}
where $R_\text{p}$ is the planet radius, $r$ the radial coordinate, $\alpha$ the coefficient of thermal expansion, and $\Delta T(r)$ the laterally-averaged temperature variation over a given time interval. \add{In addition, internal differentiation can also contribute to global variations in the planetary radius. On the one hand, melting associated with the production of secondary crust can cause radial expansion because of the lower density of the residual mantle material depleted in the incompatible elements that are enriched in the crust. On the other hand, freezing of an inner core can cause additional contraction due to the higher density of the solid part of the core that is progressively depleted in light alloying elements \citep[e.g.,][]{Grott2011,Tosi2013}.}

The interpretation of compressive tectonic features in terms of global planetary contraction is well established. It consists in mapping such features over the entire surface and using displacement–length scaling properties of faults to estimate the amount of radial shortening experienced by a given body \citep[e.g.][]{Watters2004,Byrne2014,Nahm2011,Watters2010}. With a surface dominated by compressive tectonic landforms \citeg{Strom75}---lobate scarps and wrinkle ridges---Mercury is the body for which this approach has been used most extensively. Current estimates based on  MESSENGER images suggest an overall contraction of up to 7 km \citep{Byrne2014}, with up to two additional kilometres that could be accommodated by elastic deformation prior to the formation of the observed faults \citep{Klimczak2015}. Furthermore, cross-cutting relationships between faults and impact craters indicate that global contraction on Mercury started early, \addd{about} 3.9 Ga, and continued until present at a decreasing rate \citep{Crane2017}. Models of the thermal evolution of Mercury typically predict a short initial phase of mantle heating and expansion lasting $\sim$1 Gyr, followed by cooling leading to a cumulative  amount of radial contraction that compares well with the available observations, although the onset time of contraction tends to be  slightly overestimated and the contraction rate to be constant rather than declining  \citeg{Tosi2013,Knibbe2018,Hauck2018}.  
 
Similar compilations of compressional features at a global scale have been obtained for Mars \citeg{Knapmeyer2006}. When considering all contractional structures regardless of their age, \citet{Nahm2011} inferred a maximum radial shortening of up to 3.77 km, although, as for Mercury, the expected rate of contraction would not be constant with time when taking into account the ages of the tectonic features. As discussed by \citet{Nahm2011}, the overall cooling predicted by thermal evolution models of Mars \citeg{Andrews-Hanna2008b} significantly overestimates the inferred amount of global contraction. However, the extensive volcanic resurfacing underwent by the planet during the Hesperian epoch \addd{(about 3.7 to 3.0 Ga)} \citeg{Greeley1991} may have buried older faults generated by contraction, which renders a direct comparison with thermal evolution models non-trivial. Indeed, in contrast to Mercury, for which global contraction is traditionally used as one of the primary observations to constrain the interior evolution of the planet, models of Mars generally rely on different sets of observables \citeg{Plesa2018,Samuel2019}.

On the Moon, the exceptional quality of the gravity experiment of the GRAIL mission revealed ancient igneous intrusions, indicative of an early phase of expansion of few km \citep{Andrews-Hanna2013,Elkins-Tanton2014}. Apart from tectonic structures observed within the lunar maria and associated with loading of mare basalts, compressional lobate scarps are the most common features present on the farside. Their length and surface relief, however, are much smaller than those of Mercury and Mars, implying an amount of global contraction of less than 1 km \citeg{Watters2010b,Watters2010}. Early (and strongly simplified) models starting from a cold, accretion-like interior underlying a shallow magma ocean could meet this tight constraint \citeg{Solomon1976}. More recent and sophisticated ones based on 3D thermal convection 
predict at least 2--3 km of shortening after an initial expansion phase \citep{Zhang2013b,Rolf2016}, which could be marginally compatible with the observations if additional contraction---up to 1.4 km according to \citet{Klimczak2015}---was accumulated elastically without surface manifestation.





\subsection{Heat flux and elastic lithosphere thickness}

The heat flowing through the surface from the interior provides the most direct way to probe the thermal state of a planetary body. However, heat flux measurements on an extraterrestrial body currently exist only for the Moon \citep{Langseth1976}. Taken as part of the Apollo 15 and 17 missions, these measurements were conducted near the margins of the anomalous PKT region (Section \ref{Sec__Crust}). Even if they cannot be considered to be representative of the average lunar heat flux, they can still be used as a direct constraint for the global evolution of the interior \citep{Laneuville2013,Laneuville2018}.

As part of the InSight mission to Mars \citep{Banerdt2017}, the Heat Flow and Physical Properties Package (HP$^3$), consisting of a self-hammering probe designed to measure the thermal conductivity and temperature gradient---hence the heat flux---across up to 5 m of Martian ground, was deployed on the surface of Mars on February 2019. As of July 2019, HP$^3$ has reached a depth of about 30 cm, still a small fraction of the minimum of 3 m necessary to perform a reliable heat flux estimate. 

In the absence of direct measurements, the heat flux at a specific location can be inferred by estimating the effective thickness of the elastic lithosphere ($T_\text{e}$) using gravity and topography data. Topographic features associated with observed gravity anomalies can either be compensated and hence in isostatic equilibrium in the presence of a strengthless lithosphere \citeg{Padovan2015}, or supported, partly or fully, by a lithosphere of finite strength. Although the crust-lithosphere system is characterized by a complex rheology,
its flexural behavior can be approximated with that of an elastic plate overlying a strengthless interior. By making (rather strong) assumptions on the rheological properties of the crust and lithosphere and under the approximation of small plate curvature, $T_\text{e}$ can be derived \citep{McNutt1984} and further identified with the depth of a characteristic isotherm, thereby providing clues on the thermal gradient at specific locations. Furthermore, if the age of the topographic feature responsible for lithospheric flexure is also known, for example through crater counting, an attempt can be made to reconstruct the cooling history of the mantle by estimating $T_\text{e}$ for features of different ages. Upon mantle cooling, the depth of the isotherm corresponding to $T_\text{e}$ will increase and so the elastic thickness, thus indicating a heat flux that progressively decreases over the evolution.


In addition to the above inferences based on the use of gravity and topography data, the thickness of the elastic lithosphere beneath impact basins can be determined from loading and flexure models constrained by the distribution of tectonic structures within the basin \citeg{Comer1979,Solomon1980}, or from models of viscous or viscoelastic relaxation of the basin's topographic relief \citeg{Solomon1982,Mohit2006,Kamata2015}, two approaches that have been extensively applied to the Moon. 

For Mercury, only few estimates of the elastic thickness pertaining the early evolution are available \citep{Melosh88,Nimmo04,Tosi2015,James2016}. However, these are highly discordant, with values ranging from $\sim 30-35$ km \citep{Nimmo04,James2016} to $\sim 80-100$ km \citep{Melosh88,Tosi2015} around $3.9-3.8$ Ga. The reasons of the discrepancy are not fully clear \citep{Phillips2018}, but possibly due to the different methods employed and approximations used to relate $T_\text{e}$ with the internal temperature. 

For the Moon, a variety of estimates have been made, mostly associated with impact basins. Values of $T_\text{e}$ of less than 30 km have been suggested for basins older than 3.9 Ga \citeg{Kamata2015}, possibly as small as 12 km \citep{Crosby2005}, suggesting an initially thin lithosphere and hot interior as expected for a mantle that crystallized from a magma ocean.
For younger basins such as Crisium and Imbrium, a $T_\text{e}$ of up to 75 km has been proposed \citep{Solomon1980}, in line with the idea of rapidly thickening lithosphere in a cooling mantle.

From the analysis of gravity and topography associated with a variety of structures across Mars' surface \citep[see e.g.,][for a thorough discussion of the evolution of $T_\text{e}$ 
on Mars]{Grott2013}, a consistent scenario emerges with $T_\text{e}<25$ km during the Noachian epoch
\addd{(4.5 to 3.7 Ga)}, between $\sim$50 and 150 km during the Hesperian \addd{(3.7 to 3.0 Ga)} and Amazonian \addd{(age $\leq 3$ Ga)}, though with very large temporal uncertainties, and as large as 300 km at present day beneath the northern polar cap \citep{Phillips2008}. Recently, \citet{Plesa2018} presented 3D models of Mars' interior evolution using, among others, constraints on the elastic thickness. These models agree generally well with the above estimates. In particular, the requirement of a large $T_\text{e}$ at present-day beneath the north pole poses a tight constraint, limiting the set of admissible models to those characterized by a mantle highly depleted in HPEs. In fact, a limited amount of heat sources causes the mantle to cool over the evolution to generate a sufficiently large lithosphere thickness at present.

\section{Summary and outlook} \label{Section__Summary}

Combining the various constraints and models described in the previous sections, it is possible to outline likely scenarios for the evolution of Mercury, the Moon, and Mars. Of course the picture is far from being complete, and this outlook provides a summary and indicates potential avenues for future improvements of our understanding of the evolution of stagnant-lid bodies. 

All three bodies likely had a magma ocean phase followed by fractional crystallization of the mantle, which set the stage for the subsequent evolution. The primary crust on the Moon and its traces on Mercury provide direct support to this hypothesis. A rapid formation within the first few million years of the solar system and isotopic evidence from meteorites favor a magma ocean also for Mars. A global-scale overturn of the solid mantle---beneath the stagnant lid or involving the lid---due to the gravitational instability of \addd{FeO}-rich cumulates is a direct consequence of fractional crystallization and the most widely explored scenario. However, for a slowly solidifying magma ocean, as is possibly the case for Mars and most likely for  the Moon, convection and mixing can start during solidification. Existing models of the long-term evolution of these bodies constrained by observations mostly rely on the assumption of a homogeneous mantle. The long-term consequences of the complex processes associated with magma ocean crystallization are receiving more and more attention but are yet to be explored in detail for each planet.

\begin{figure}[ht!]
\centering
\includegraphics[width=0.5\textwidth]{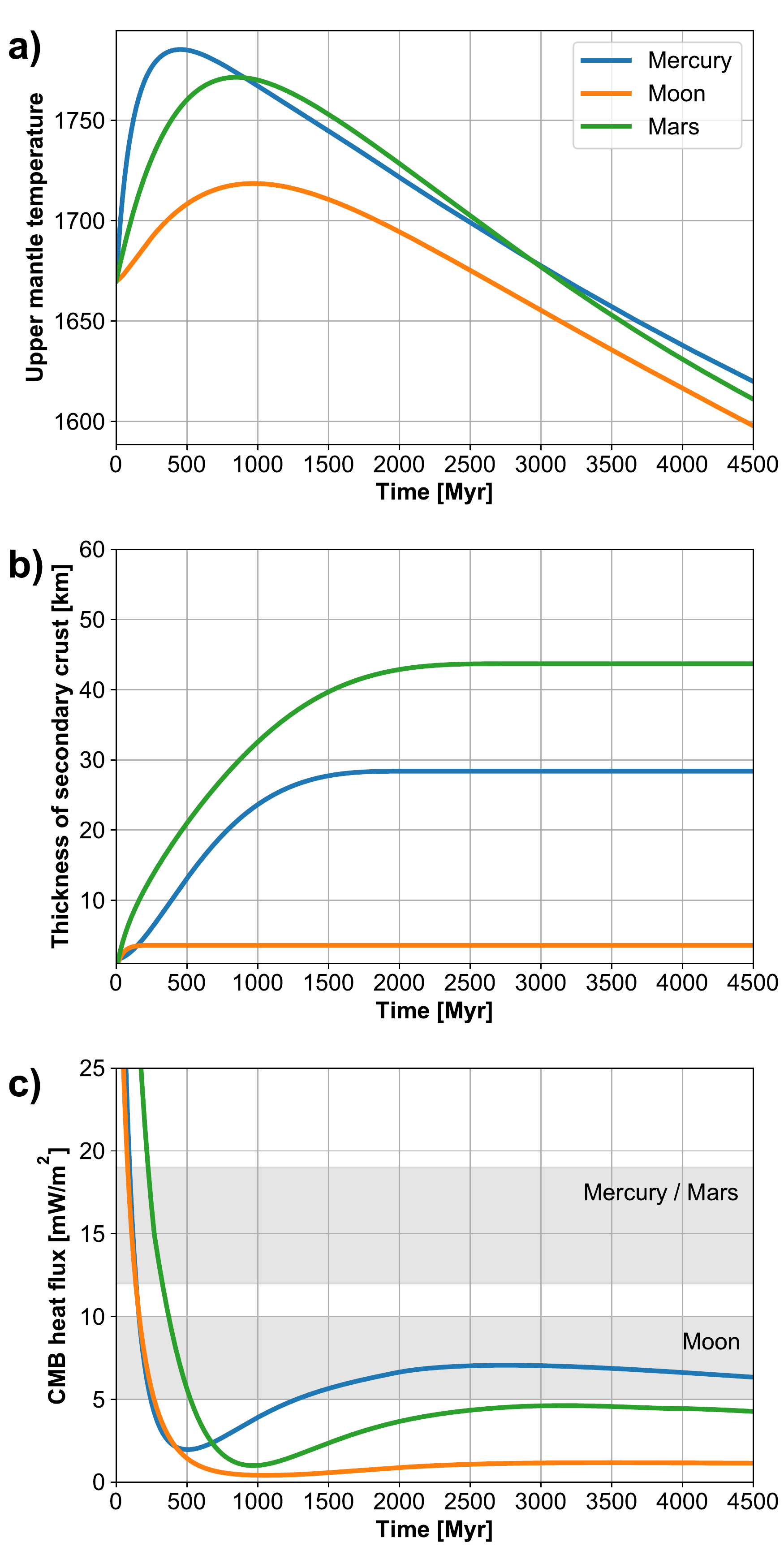}
\caption{\add{Possible} evolution of the a) upper mantle temperature (i.e., the temperature beneath the stagnant lid), b) \add{thickness of the} secondary crust (which would sum up to an existing primary crust), and c) heat flux at the CMB for Mercury (blue lines), the Moon (orange lines), and Mars (green lines). In panel c, the two grey areas indicate a possible range of core adiabatic heat flux for Mercury and Mars (upper area between 12 and 19 mW/m$^2$), and for the Moon (lower area between 5 and 10 mW/m$^2$). The curves are obtained using parameterized thermal evolution models similar to those of \citet{Grott2013} and \citet{Tosi2013}, to which we refer for a complete description. Model parameters for Mercury are taken from  \citet{Hauck2018}, for Mars from \citet{Grott2013}, and for the Moon from \citet{Laneuville2018}. }
\label{fig:summary}
\end{figure}

The presence of the stagnant lid early on in the evolution (possibly following a brief episode of surface mobilization at the end of the magma ocean phase) causes inefficient heat loss, and hence heating and expansion of the mantle (Figure \ref{fig:summary}a), and crust formation (Figure \ref{fig:summary}b). The segregation of HPEs due to the formation of the primary crust of the Moon left the mantle largely depleted, which strongly limited the production of large volumes of secondary crust, but not of volcanism in general, as the lunar maria clearly show. The crusts of Mercury and Mars are instead mostly volcanic (i.e., secondary) and grew to the thickness inferred at present day from gravity and topography data within the first few hundred million years of evolution due to the production of partial melt, implying that a substantial amount of HPEs remained available to drive mantle convection and decompression melting after magma ocean solidification.

The phase of mantle heating and crust production is followed by secular cooling (Figure \ref{fig:summary}a), radial contraction, and by the increase of the thickness of the lithosphere and its elastic part. Surficial evidence is provided by the extent and number of compressive tectonic features, which allow the amount of radial contraction accumulated through a planet's history to be estimated. Such features are present on all the three bodies, particularly on Mercury, whose estimated contraction of about 7 km can be matched by evolution models. 
The smaller global contraction of Mars ($<4$ km) and the Moon ($<1$ km) are generally difficult to match by evolution models, which instead predict larger values, thus suggesting that part of the accumulated contraction may have been stored elastically with no surface expression. The effective elastic lithosphere thickness, estimated at specific locations by combining gravity and topography data with flexure modelling, increases with time due to mantle cooling. This behavior has been mostly recognized for Mars and is confirmed by numerical models. Large uncertainties associated with dating of the corresponding surface units somehow limit the use of this constraint.

In the early phases of the solar system
many large objects collided with the 
newly formed planets, and the history
of this tumultuous past is manifest
in the battered surfaces of Mercury, the
Moon, and to a somewhat lesser extent, Mars.
The energies involved in this kind of 
events are extremely large, enough 
to interact with mantle convection,
thus affecting the overall evolution,
with melting activity, thus affecting the
secondary crust production, and possibly 
with the core heat flux, thus affecting dynamo
action.
The most energetic events occur
early on, and their dating is often
quite uncertain, making the
connection between these exogenous
events and the interior processes
they may interact with not 
straightforward to assess. 
However, it is clear that the
mutual interaction depends on 
the interior thermal state,
and, as such, large impact basins
provide a window into the 
interior processes at the time 
of impact.
There is evidence that the
formation of the several 
large Martian basins did temporally
stop dynamo action in the mantle,
and that the properties of
large lunar basins confirm the 
existence of large scale compositional
heterogeneities in the mantle
of the Moon. 
Compared to Mars and the Moon,
data for Mercury are relatively sparse. 
However, as in the case of the
Moon, the properties of its
large basins are compatible with 
the overall evolution of the 
planet, as inferred from the
entire set of available constraints
(Figure \ref{fig:summary}).

Although at present-day only Mercury possesses a global magnetic field, an early one was active on all the three bodies, as the magnetization of their crust and lithosphere demonstrates (Figure \ref{fig:magfield}). Thermal dynamos driven by a super-adiabatic heat flux extracted from the core by the convecting mantle are short-lived (Figure \ref{fig:summary}c). A thermal dynamo on Mercury would have likely stopped before 3.7 Ga (the age of the observed magnetized region). Furthermore, it is not known whether or not Mercury's magnetic field operated continuously from this age until present. Evolution models considering a core with Fe-Si composition and accounting for the effects of its crystallization seem to suggest so. An early episode of strong mantle and core cooling due to surface mobilization or to an initially hot core may have been responsible for the Martian dynamo, which ceased to operate $\sim$4 Ga. This mechanism is not sufficient to power the lunar dynamo, which survived until 2.5 Ga, possibly much longer, and generated $\sim$4 Ga a magnetic field of high intensity, which no single mechanism can account for.

\add{The global evolutions of Mercury, the Moon and Mars illustrated in Figure \ref{fig:summary} are valid for a specific model with a specific set of parameters. Yet,} the figure shows that overall the evolution of the three bodies, as constrained
by the observations described in this chapter,
which are different for each body,
is largely similar.
Thus, one could surmise that thermal evolution models
do capture the most important physical processes relevant
to the thermo-chemical evolution of stagnant-lid bodies.
However, given the large number of parameters that
are required to run these kind of models (typically
in excess of about 20), it is fair to admit
that there is often enough room to adjust 
them in order to match the observations.
A better leverage on the predictive and 
inference powers of these models  
would require major advances 
along two main avenues: 
material properties and 
observational constraints.
In terms of material properties
we take viscosity as the most
relevant example, since it represents 
a parameter that has a huge
effect on the thermal evolution, but 
for which only very broad 
constraints exist. 
Viscosity is difficult 
to measure both in  the laboratory 
and with theoretical calculations.
Furthermore, in a planetary setting, 
it is affected by several poorly 
known and difficult to constrain 
factors, such as presence of melt, 
presence of volatiles, \add{grain size,} and dominant creep
mechanism.
These unavoidable obstacles imply 
that improvements in 
observational constraints are the most
likely way to refine our understanding
of the evolution of stagnant-lid bodies.
With their potential ability of
inferring interior structure, heat flux,
crustal structure, and possibly density and
temperature structures of the mantle,
geophysical stations, such as the 
InSight lander currently located 
on the surface of Mars, 
are poised to provide a big
step forward in our understanding
of the detailed evolution history of
the stagnant-lid bodies of the 
inner solar system.
Complementary refinements of the
dating of the surfacial units, possible
by direct analysis of samples, which
will be facilitated by future sample
return missions,
will also improve the information
content of a large set of additional
constraints, from large basins properties
to lid thickness estimates, which 
critically depend on the age of
the associated crustal units.


It is somewhat self-evident 
that the level of complexity
of a given model must depend 
on the detail, amount, and variety of the 
observations that the model is 
trying to explain.
Future missions will provide new data,
and evolution models will obviously 
need to be expanded and 
possibly adapted and tailored
specifically 
for the different bodies.
This future development will
necessarily make the
straightforward and
informative comparison
of the evolution of the 
different bodies, as 
presented in Figure \ref{fig:summary},
not possible. 
But this will only indicate that
the field will have matured, 
which is only to be welcomed, 
as long as modelers will not attempt
to reach the level of exactitude 
of the ancient Art of Cartography
\citep{Borges1999}.

\section*{Acknowledgments}
We thank Maxim Ballmer for inviting us to write this manuscript and two anonymous reviewers for their comments that helped improve it. We also thank Maxime Maurice for his assistance in setting up the model used to generate Figure 1. N.T. acknowledges support by the Helmholtz Association (project VH-NG-1017) and S.P. by the DFG (Research Unit FOR 2440 Matter under planetary interior conditions).



\newpage
\bibliographystyle{myplainnat2}

\end{document}